\documentclass[preprint]{aastex}

\newcommand{\be}{\begin{equation}}
\newcommand{\ee}{\end{equation}} 
\newcommand{\crcon}{ {\cal C}_r } 
\newcommand{\mout}{{ m_2 }} 
\newcommand{\diffuse}{{ \mathcal{D} }} 
\newcommand{\erf}{{ {\rm Erf} }} 
\newcommand{\ebar}{{ \langle e \rangle }}

\newcommand{\etarms}{{ \eta_{\rm rms} }} 
 
\newcommand{\norb}{{ N_{\rm orb} }} 
\newcommand{\nens}{{ N_{\rm ens} }}

\newcommand{\tvar}{{ T }} 
\newcommand{\pnew}{{ {Q} }} 
\newcommand{\cfun}{ \mathcal{C} } 
\newcommand{\cfuntwo}{ \mathcal{C}^{(2)} } 
\newcommand{\tdiff}{{ t_{\rm diffuse} }}
\newcommand{\texp}{{ t_{\rm exp} }}
\newcommand{\pbd}{{ P_{\, {\rm b}} }}

\begin{document} 

\title{MEAN MOTION RESONANCES IN EXTRASOLAR PLANETARY SYSTEMS WITH TURBULENCE, INTERACTIONS, AND DAMPING} 

\author{Daniel Lecoanet\altaffilmark{1,2}, Fred C. Adams\altaffilmark{1,3}, 
and Anthony M. Bloch\altaffilmark{1,4} } 

\altaffiltext{1}{Michigan Center for Theoretical Physics \\
Physics Department, University of Michigan, Ann Arbor, MI 48109} 

\altaffiltext{2}{Physics Department, University of Wisconsin, Madison, WI 53706} 

\altaffiltext{3}{Astronomy Department, University of Michigan, Ann Arbor, MI 48109} 

\altaffiltext{4}{Department of Mathematics, University of Michigan, 
Ann Arbor, MI 48109}

\begin{abstract} 

This paper continues our previous exploration of the effects of
turbulence on mean motion resonances in extrasolar planetary systems.
Turbulence is expected to be present in the circumstellar disks that
give rise to planets, and these fluctuations act to compromise
resonant configurations.  This paper extends previous work by
considering how interactions between the planets and possible damping
effects imposed by the disk affect the outcomes. These physical
processes are studied using three related approaches: direct numerical
integrations of the 3-body problem with additional forcing due to
turbulence, model equations that reduce the problem to stochastically
driven oscillators, and Fokker-Planck equations that describe the time
evolution of an ensemble of such systems. With this combined approach,
we elucidate the basic physics of how turbulence can remove extrasolar
planetary systems from mean motion resonance. As expected, systems
with sufficiently large damping (dissipation) can maintain resonance,
in spite of turbulent forcing. In the absence of strong damping,
ensembles of these systems exhibit two regimes of behavior, where the
fraction of the bound states decreases as a power-law or as an
exponential. Both types of behavior can be understood through the
model developed herein.  For systems that have weak interactions
between the planets, the model reduces to that of a stochastic
pendulum, and the fraction of bound states decreases as a power-law
$\pbd \propto t^{-1/2}$.  For highly interactive systems, however, the
dynamics are more complicated and the fraction of bound states
decreases exponentially with time. We show how planetary interactions
lead to drift terms in the Fokker-Planck equation and account for this
exponential behavior.  In addition to clarifying the physical
processes involved, this paper strengthens our original finding that
turbulence implies that mean motions resonances should be rare.

\end{abstract} 


\section{INTRODUCTION}  

A sizable fraction of the observed extrasolar planetary systems with
multiple planets have period ratios that are close to the ratios of
small integers, and hence are candidates for being in mean motion
resonance (e.g., Mayor et al. 2001, Marcy et al. 2002, Butler et
al. 2006).  However, a true resonance not only requires particular
period ratios, but also oscillatory behavior of the resonant angles,
which in turn requires specific values for the orbital elements of the
two planets (Murray \& Dermott 1999; hereafter MD99).  As a result, a
mean motion resonance represents a rather special dynamical state of
the system.  The suspected origin of such configurations is through a
process of convergent migration, wherein two planets are moved inward
together through their solar system by a circumstellar disk and
thereby given the opportunity to enter into a resonant state (e.g.,
Lee \& Peale 2002, Lee 2004, Beaug{\'e} et al. 2006, Moorhead \& Adams
2005, Crida et al. 2008).  Since these disks are likely to be
turbulent (e.g., Balbus \& Hawley 1991), and since resonant states are
easily disrupted, it is possible for turbulent fluctuations to drive
planetary systems out of mean motion resonance.  In an earlier paper
(Adams et al. 2008; hereafter Paper I), we explored this possibility
and found that the presence of turbulence implies that mean motion
resonances should be rare. This paper continues this earlier effort by
exploring the problem in greater depth and by including additional
physical effects.

This work is motivated by the current observations of extrasolar
planetary systems with multiple planets.  Although many of the
observed systems display period ratios that are consistent with mean
motion resonance, more data is required to determine if many of these
systems are truly in resonance.  The review of Udry et al. (2007)
shows that four systems have 2:1 period ratios, 2 systems have 3:1
period ratios, 2 systems have 4:1 period ratios, and 4 systems have
5:1 period ratios. Of these possible resonant systems, those with 2:1
period ratios are the most compelling candidates, with ratios
$P_2/P_1$ = $2 \pm 0.01$. Of these systems, one case is GJ876, which
is well observed and has two planets thought to be deep in a 2:1 mean
motion resonance (Marcy et al. 2001, Laughlin \& Chambers 2001). The
other candidates for systems in 2:1 resonance include HD 82943
(Gozdziewski \& Maciejewski 2001, Mayor et al. 2004, Lee et al. 2006),
HD 128311 (Vogt et al.  2005), and HD 73526 (Tinney et al. 2006,
S{\'a}ndor et al.  2007).  The status of these latter three systems
remain unresolved; in any case, their libration widths are wide (as
well as uncertain) and hence these systems are not as ``deep'' in
resonance as the GJ876 system.  In addition, the 55 Cancri system
(Marcy et al. 2002) has planets with period ratios close to 3:1, and
has been suggested as another candidate for mean motion resonance (Ji
et al. 2003); however, subsequent work (Fischer et al.  2008)
indicates that the system is unlikely to be in resonance.  Finally, we
note that a new system of ``Super-Earths'' (with masses of 4.2, 6.9
and 9.2 $M_E$) has recently been discovered orbiting HD 40307 with
periods of 4.3, 9.6, and 20.5 days (Mayor et al. 2008). This system
thus has period ratios that are relatively close to 4:2:1, but are
most likely too distant to be in resonance. Although more data is
required, the observational landscape can be roughly summarized as
follows: one system (GJ876) is deep in a 2:1 mean motion resonance,
three more systems are either in resonance or relatively close, and
many more planetary systems have period ratios indicative of mean
motion resonance but are probably not actually bound in resonant
states.  Since mean motion resonances are relatively easy to
compromise, this collection of observed systems provides important
clues regarding their formation and past evolution.

In Paper I, we considered turbulent fluctuations as a mechanism to
remove systems from mean motion resonances while retaining
nearly-integer period ratios. If the fluctuations have sufficiently
large amplitude and duty cycle, turbulence is indeed effective at
removing systems from bound resonant states. Specifically, this
earlier work used both a simple model equation (a stochastic pendulum)
and direct integrations based on the architecture of the observed
planetary system GJ876 (see above).  In both cases, turbulent forcing,
with the amplitudes expected from magnetohydrodynamic (MHD)
simulations (e.g., Nelson \& Papaloizou 2003; Laughlin et al. 2004,
hereafter LSA04; Nelson 2005) of magneto-rotational instability (MRI),
were found to drive systems out of a resonant state. We note that not
all circumstellar disks will support MRI, especially when they are
sufficiently resistive. Nonetheless, this treatment applies to any
type of stochastic fluctuations, including all types of turbulence,
that could be present in such disks.  Although our earlier work
elucidates the basic physics of the problem, and shows that turbulence
can be effective at removing systems from resonance (Paper I), several
issues must be studied further:

[1] The planets migrate inward during much of the time while
turbulence is expected to be present. As a result, the effects of
turbulence on mean motion resonance during the migration epoch must be
considered. This issue has been considered in recent work (Moorhead
2008), which shows that turbulence that acts during the epoch of
planet migration does indeed compromise mean motion resonance in a
manner that is qualitatively similar to the results of Paper I; a more
detailed study of this process is underway (Moorhead \& Adams, in
preparation).

[2] The circumstellar disk that produces both the inward migration and
the turbulent fluctuations can also produce a net damping of the
eccentricity of the planetary orbits. This damping, in turn, can act
as a damping mechanism to counter the action of turbulence driving
systems from resonance. The process of convergent migration itself,
which allows planets to enter into mean motion resonance (e.g., Lee \&
Peale 2002; Crida et al. 2008), can also provide a damping mechanism
for the resonance. More generally, damping can effectively take place
whenever the fluctuations have a nonzero first moment (R. Malhotra,
private communication) or through dynamical viscosity (E. Zweibel,
private communication).  In any case, the combined effects of damping
and turbulent fluctuations on mean motion resonance should be studied
further, and are considered here (see \S 3).

[3] The stochastic pendulum model (used in Paper I) reduces the full
physical system --- which has 18 phase space variables for a two
planet solar system --- to a much simpler system with only one
variable. In its simplified form, the system can freely random walk
both in and out of resonance through the action of the turbulent
fluctuations. As mentioned above, however, a true resonance represents
a special dynamical state of the system, and hence the full system
(with 18 phase space variables) will not necessarily re-enter
resonance as easily as the one variable system. In Paper I, we found
that the full system can in fact random walk back into resonance after
leaving, but not as easily as suggested by the stochastic pendulum
model. As a result, the issue of how a planetary system re-enters into
resonance after leaving, especially in the presence of fluctuations,
remains open. Any barrier to re-entry into resonance causes the
fraction of bound states (systems in resonance) to decrease with time
faster than the stochastic pendulum model (see \S 4). A related issue 
is that highly interactive systems, those with large planets and 
high eccentricities, tend to leave resonance more readily than systems 
that are less interactive. 

[4] When planets are in resonance, they are protected (at least in
part) from orbit crossings, even when the eccentricities are fairly
large. When the planets are knocked out of resonance, however, they
can be subject to greater eccentricity excitation and eventual orbit
crossing, and hence have a chance to scatter off each other and be
ejected.  As a result, solar systems that leave resonance have a
greater chance of losing planets. Further, when a system ejects a
planet, that system can never re-enter a resonant state; this loss of
planets contributes to the decreasing number of bound states (systems
in resonance) as a function of time. We also expect highly interactive
systems to eject planets more readily than those that are less
interactive.  The role of planet ejection on the time evolution of the
fraction of bound states must be explored further.

The goal of this paper is to study the effects of turbulence on mean
motion resonance with the above issues in mind. In \S 2, we review the
three approaches to the problem used herein: The basic model of the
resonant system as a (stochastic) pendulum, the corresponding phase
space approach to the dynamics using a Fokker-Planck equation, and
full numerical integrations (using all 18 phase space variables for
the 3-body problem). This section also presents new ensembles of
numerical simulations to illustrate how well the systems are described
by the stochastic pendulum model and where its shortcomings are
found. In \S 3 we consider the combined effects of damping and
turbulence on mean motion resonance.  In \S 4 we derive a more
detailed model of mean motion resonance, where we keep higher order
terms that include effects from planet-planet interactions.  Finally,
we conclude in \S 5 with a summary of our results and a discussion of
their implications.

\section{FORMULATION} 

In this section we outline the three basic methods used in this paper.
We first review the simplest version of the pendulum model for mean
motion resonance (from MD99; see \S 2.1) and then outline our
treatment for including turbulent fluctuations (from Paper I; see \S
2.2). Our first approach is to directly integrate the resulting
stochastic differential equations that describe the time evolution of
resonances in the presence of turbulence; in this case, a large
ensemble of different solutions must be explored to sample the effects
of different realizations of the stochastic fluctuations. An alternate
approach, outlined in \S 2.3, is to find the corresponding
Fokker-Planck equation, which describes the time evolution of the
distribution of states. Our third and final technique is to directly
integrate (numerically) the full 3-body system including velocity
perturbations to account for the turbulence (\S 2.4); this approach
also requires a large ensemble of different realizations of the
problem to build up a statistical description.

\subsection{The Basic Pendulum Model} 

As a starting point, we use a version of the formalism presented in
MD99.  To start, this treatment is limited to the circular restricted
3-body problem and hence considers the libration angle $\phi$ for two
planets in resonance to have the form
\be
\phi = j_1 \lambda_2 + j_2 \lambda + j_4 \varpi \, , 
\label{phidef} 
\ee
where $\lambda$ and $\lambda_2$ are the mean longitudes of the two planets 
and $\varpi$ is the longitude of periapse of the inner planet.  In this 
case, the equation of motion for the resonance angle $\phi$ reduces to
that of a pendulum, i.e.,
\be 
{d^2 \phi \over dt^2} + \omega_0^2 \sin \phi = 0 \, . 
\label{pendone} 
\ee 
The natural oscillation frequency $\omega_0$ of the pendulum --- 
the libration frequency of the resonance angle --- is given by 
\be 
\omega_0^2 = - 3 j_2^2 \crcon \Omega e^{|j_4|} \, , 
\qquad {\rm where} \qquad \crcon = \mu \Omega \, \alpha f_d(\alpha) \, . 
\label{omega} 
\ee
Here, $\Omega$ and $e$ are the mean motion and eccentricity of the
inner planet, and ($j_2$, $j_4$) are integers that depend on the type
of resonance being considered. The parameters $\crcon$, $e$, and
$\Omega$ are assumed to be constant for purposes of determining the
frequency. The mass ratio $\mu$ = $\mout/M_\ast$, where $\mout$ is the
mass of the outer planet and $M_\ast$ is the stellar mass.  The
quantity $\alpha f_d (\alpha)$ results from the expansion of the
disturbing function (MD99), and the parameter $\alpha$ is the ratio of
the semimajor axes of the two planets, i.e., $\alpha \equiv
a_1/a_2$. Note that this approximation scheme (adapted from MD99)
neglects terms of order ${\cal O}(\mu)$; we consider higher order
terms in \S 4.

Many of the observed (candidate) resonant systems are in or near the
2:1 mean motion resonance; for the sake of definiteness, we focus on
that case. This resonance is also generally the strongest.  For the
2:1 mean motion resonance, the integer $j_2 = -1$ and $\alpha
f_d(\alpha) \approx -3/4$ (MD99). In this case, the natural
oscillation frequency $\omega_0$ of the libration angle is given by
\be
\omega_0^2 \approx {9 \over 4} \mu e^{|j_4|} \, \Omega^2 \,  . 
\label{omegadef} 
\ee 
Typical planet masses in observed extrasolar planetary systems are of
order a Jovian mass so that $\mu \sim 10^{-3}$.  The eccentricity can
vary over a wide range, with $e \sim 0.1$ in order of magnitude, but
the median eccentricity is closer to $e \sim 0.3$. The relevant 
values of $|j_4|$ = (1,2), and we generally use $|j_4|$ = 2 (see the 
discussion of \S 2.4).  As a result, the ratio of frequencies
$\omega_0 / \Omega \sim 10^{-2}$ and hence the period of the libration
angle is expected to be $\sim100$ orbits. With this frequency
specified, we define a dimensionless time variable
\be
\tau = \omega_0 t \, , 
\ee 
and write the pendulum equation in dimensionless form 
\be 
{d^2 \phi \over d\tau^2} + \sin \phi = 0 \, ,
\label{pendtwo} 
\ee 

Note that the energy scale associated with the potential well of the
resonance, modeled here as a pendulum, is much smaller than the
binding energy of the planets in their orbits. Neglecting
dimensionless numbers of order unity, we can write the specific energy
$E_p$ of the pendulum as 
\be 
E_p \sim \omega_0^2 a^2 \sim \mu e^{|j_4|} \, {G M_\ast \over a} 
\sim \mu e^{|j_4|} \, E_{\rm orb} \sim 10^{-4} \, E_{\rm orb} \, ,
\label{resenergy} 
\ee
where $E_{\rm orb}$ is the energy (per unit mass) of the planet in the
gravitational potential well of the star. As result, we estimate that 
$E_p / E_{\rm orb} = {\cal O} (\mu e^{|j_4|})$, so that the potential 
well of the resonance is about $10^4$ times shallower than the
potential well of the star. As a result, planets can be removed from
resonance much more easily than they can be removed from orbit (by
turbulent fluctuations --- note that planets are often ejected by
scattering after orbit crossing).

\subsection{Turbulent Fluctuations} 

The net effect of turbulence is to provide stochastic forcing
perturbations on the mean motion resonance, which is modeled here as a
pendulum. To include this stochastic process, the equation of motion
(\ref{pendtwo}) can thus be modified to take the form 
\be
{d^2\phi \over d \tau^2} + \left[ 1 + \eta_k\delta([\tau]-\Delta\tau) \right] 
\sin\phi = 0 \, , 
\label{pendturb} 
\ee
where $\eta_k$ sets the amplitude of the turbulent forcing and $\Delta
\tau$ sets the time interval. These quantities are derived in Paper I 
(including a discussion of where to introduce the turbulent term) and 
are discussed further below.  

Briefly, the time scale $\Delta \tau$ is the time (in dimensionless
units) required for the turbulence in the disk to produce an
independent realization of the fluctuations.  This time scale is
typically one or two orbit times at the disk location in question
(LSA04, Nelson 2005). Since the relevant part of the disk extends
outside the orbit of the outer planet, and since the outer planet has
twice the period of the inner planet, the time scale $\Delta \tau
\approx 4 \omega_0 (2 \pi / \Omega) = 8 \pi \omega_0/\Omega$.

Next we need to specify the forcing strength over the time interval
$\Delta \tau$.  In the original derivation of the pendulum equation
(MD99), one finds
\be
{d^2 \phi \over d t^2} \sim {d \Omega \over dt} \sim - {\Omega \over 3} 
\left( {1 \over J} {dJ \over dt} \right) \, , 
\ee 
where the second approximate equality follows from the relationship
between the mean motion and the orbital angular momentum. Converting
to dimensionless form and integrating over one cycle to produce a
discrete quantity, one finds 
\be
\eta_k \approx - {1 \over 3} {\Omega \over \omega_0} 
\left( {\Delta J \over J} \right)_k \, . 
\ee
The amplitudes $[(\Delta J)/J]_k$ have been calculated for a variety
of cases using MHD simulations (e.g., Nelson \& Papaloizou 2003, 2004;
LSA04; Nelson 2005). In basic terms, the torque exerted on a planet by
the disk will be a fraction of the benchmark scale $T_D = 2 \pi G
\Sigma r m_P$, where $\Sigma$ is the disk surface density (Johnson et
al. 2006). The amplitude for angular momentum variations is thus given
by $\Delta J = f_T \Gamma_R T_D (8 \pi/\Omega)$, where the factor 
$8 \pi/\Omega$ is the time over which one realization of the
turbulence acts, $f_T \sim 0.05$ is the fraction of the benchmark
scale $T_D$ that applies for a disk without a planet, and $\Gamma_R
\sim 0.1$ is the reduction factor due to the production of a gap in
the disk (and hence loss of disk material).  Including all of these
factors (see Paper I), the relative fluctuation amplitude is given by
$[(\Delta J)/J]_{\rm rms} \sim 10^{-4}$ under typical conditions.
With this typical amplitude for $[(\Delta J)/J]_k$ and $\Omega/\omega_0
\sim 100$ (see above), we expect that $\etarms \sim 0.005$.

In general, the fluctuation amplitude will vary from case to case over
a range of at least an order of magnitude due to different levels of
turbulence and varying surface densities, or equivalently, varying
disk masses. The amplitude quoted above is applicable for planets of
Jovian mass. Lighter planets produce smaller gaps and have larger
values of the factor $\Gamma_R$ and hence larger forcing amplitudes
$[(\Delta J)/J]_k$. Note that even planets that are too small to clear
a gap will still have an effective value of $\Gamma_R < 1$ because the
wakes produced by the planet compromise the turbulent fluctuations in
the vicinity (Nelson \& Papaloizou 2004). In addition, different disk
systems will retain their disk mass for a range of lifetimes, and this
effect leads to another source of system to system variation.
Finally, even for a given initial disk mass and disk lifetime, the
amount of turbulence that resonant systems experience depends on how
late (in the life of the disk) the planets are formed and/or captured
into resonance (e.g., Thommes et al. 2008). These variations thus allow
for a wide range of possible outcomes. 

For completeness, we note that turbulence is not the only possible
source of stochastic fluctuations acting on planets in mean motion
resonance. For example, a residual disk of planetesimals can provide a
granular background gravitational potential and act in a qualitatively
similar manner (Murray-Clay \& Chiang 2006). The formalism developed
here can be used with any source of stochastic fluctuations with a
given amplitude $\eta_k$ and forcing time interval $\Delta \tau =
\omega_0 (\Delta t)$.

\subsection{Phase Space Distribution Functions} 

An important technique used to analyze the behavior of stochastic
differential equations --- including those describing mean motion
resonance --- is to work in terms of phase space variables. For the
stochastic pendulum considered here, we rewrite the equation of motion
in the form 
\be 
{d \phi \over d \tau} = V \, \qquad {\rm and} \qquad {d V \over d \tau} 
= - \left[ 1 + \eta_k\delta([\tau]-\Delta\tau) \right] \sin \phi \, . 
\ee 
The probability distribution function $P(\tau, \phi, V)$ for these
phase space variables obeys the Fokker-Planck equation (e.g., Mallick
\& Marcq 2004, hereafter MM04) which takes the form 
\be 
{\partial P \over \partial \tau} + V {\partial P \over \partial \phi} 
- \sin \phi {\partial P \over \partial V} = {1 \over 2} D 
\sin^2 \phi {\partial^2 P \over \partial V^2 } \, . 
\label{fpzero} 
\ee
As outlined in Paper I, the phase space diffusion constant $D$ is
specified by the amplitude $\eta_k$ of the turbulent fluctuations 
and the time interval $\Delta \tau$ required for them to attain an 
independent realization, i.e., 
\be
D = {\langle \eta_k^2 \rangle \over \Delta \tau} \, . 
\label{diffusedef} 
\ee 
In most cases of interest, the angle $\phi$ varies more rapidly than
the velocity $V$ (see Paper I and MM04). As a result, we can average
the Fokker-Planck equation (\ref{fpzero}) over the angle $\phi$ to
obtain the corresponding time evolution equation for the
$\phi$-averaged probability distribution function $P(\tau, V)$.
Throughout this paper, we use the same symbol (here, $P$) for the
distribution function both before and after an averaging procedure;
this choice simplifies the notation, but could result in some
ambiguity, although the relevant version of the distribution function 
should be clear from the context (this issues also arises in \S 4). 
After averaging, the resulting Fokker-Planck equation (\ref{fpzero}) 
becomes a basic diffusion equation 
\be
{\partial P \over \partial \tau} = {D \over 4} 
{\partial^2 P \over \partial V^2} \, , 
\ee 
which can be solved to obtain the result 
\be
P(\tau, V) = {1 \over (\pi D \tau)^{1/2} } 
\exp \left[ - {V^2 \over D \tau} \right] \, . 
\label{diffsolve}
\ee
This solutions shows that the energy of the pendulum can grow large at
long times. In this limit, the kinetic energy dominates the potential 
energy term, and we approximate the energy using $E \approx V^2/2$. 
The resulting probability distribution function for the energy then 
can be written in the form \be
P (E, \tau) = \left( {2 \over \pi D \tau} \right)^{1/2} E^{-1/2}  
\exp \left[ - {2E \over D \tau} \right] \, . 
\label{edistrib} 
\ee
For this solution for the distribution function, the expectation value
of the energy grows linearly with time in the long time limit, i.e.,
$\langle E \rangle = D \tau/ 4$. As the mean energy increases, the
probability of the system remaining bound in a low energy (resonant)
state decreases. Here we use $E > 1$ as the requirement for the
libration angle to circulate and hence for the resonance to be
compromised (see Paper I for further discussion). The probability 
$\pbd$ of remaining bound is thus given by 
\be 
\pbd (\tau) = 
\left( {2 \over \pi D \tau} \right)^{1/2} \int_0^1 {dE \over \sqrt{E} } 
\exp \left[ - {2E \over D \tau} \right] \, = {2 \over \sqrt{\pi}} 
\int_0^{z_0} {\rm e}^{-z^2} dz \, = \erf (z_0) \, , 
\label{pbint} 
\ee
where $z_0 = (2/D \tau)^{1/2}$ and where $\erf (z)$ is the error
function (e.g., Abramowitz \& Stegun 1970). In the limit of late
times, when $z \ll 1$, the error function has the asymptotic form 
$\erf (z) \sim 2z/\sqrt{\pi}$, and the probability $\pbd$ that the
planetary system remains in resonance is given by 
\be
\pbd (\tau) \approx \left( {8 \over \pi D \tau} \right)^{1/2} = 
\left( {2 \over \pi \langle E \rangle} \right)^{1/2} \, , 
\label{probtime} 
\ee
where this expression is valid for sufficiently late times 
when $D \tau \gg 1$.  

\subsection{Numerical Integrations} 

The third and final technique that we use to study this problem is
direct numerical integration of the planetary systems. In other words,
we integrate the full set of 18 phase space variables of the
corresponding 3-body problem (using a B-S integration scheme, e.g.,
Press et al. 1992).  Turbulence is included by applying discrete
velocity perturbations at regular time intervals; for the sake of
definiteness, the forcing intervals are chosen to be twice the orbital
period of the outer planet (four times the period of the inner
planet). Both components of velocity in the plane of the orbit are
perturbed randomly, but the vertical component of velocity is not
changed. The amplitude of the velocity perturbations is then allowed
to vary.

For the numerical experiments carried out for this paper, we start a
large ensemble of $\nens$ planetary systems in the 2:1 mean motion
resonance, and then integrate forward in time using the velocity
perturbations (as described above) to model turbulent fluctuations.
Throughout this work, we use the resonance angle defined by equation
(\ref{newangle}) for interactive systems such as GJ876 (see \S 4), and
its analog in the circular, restricted 3-body problem given by
equation (\ref{phidef}) with $|j_4|$ = 2.  Our numerical experiments
show that this choice of resonance angle allows the systems to remain
in bound (resonant) states for longer times than for the standard
case where $|j_4|=1$ (see Paper I and \S 4). Since we are primarily
interested in how turbulent fluctuations can compromise mean motion
resonance, we want to consider the cases that are most robust
(longest-lived). Finally, we note that the analysis of this paper can
be performed for a host of other forms for the resonance angle.

Over the course of time, systems move both in and out of resonance.
To monitor this behavior, we determine the maximum libration amplitude
of the resonance angle over a fixed time interval.  This time scale is
chosen to be somewhat longer than the libration time, which is much
shorter than the evolution time, i.e., the time scale over which a
substantial fraction of the systems leave resonance.  Note that the
libration time is typically $\sim100$ orbits (see equation
[\ref{omegadef}]), whereas the evolution time is of order $10^6$
orbits.  For the sake of definiteness, we consider a planetary system
to be in resonance when its maximum libration angle is less that 90
degrees; for larger maximum libration angles, we consider the
resonance to be compromised. Note that the boundary at 90 degrees is
somewhat arbitrary; fortunately, however, the statistics for the
fractions of bound systems is insensitive to this value (see Paper I).

Although the observed set of extrasolar planetary systems shows an
astonishing degree of diversity, here we consider only two particular
examples that bracket the possibilities. For the first case, we
consider analogs of the GJ876 planetary system, which has two planets
that are observed to be deep in a 2:1 mean motion resonance (Marcy et
al. 2001). This system has a relatively small central star (with mass
of only $M_\ast$ = 0.32 $M_\odot$) and relatively large planets (with
masses $m$ = 0.79 and $m_2$ = 2.52 $m_J$). The eccentricities are
substantial, with $e$ = 0.26 for the inner planet and $e_2$ = 0.034
for the outer planet.  The periods are approximately 30 and 60 days,
respectively (see Rivera et al. 2005 for further detail). With these
parameters, the GJ876 system is the most highly interactive system
observed to date. Although comparable solar systems with larger
planetary masses and/or higher orbital eccentricities would be even
more interactive, they would also most likely be unstable (Laughlin \&
Chambers 2001). Since the pendulum model for mean motion resonance
represents the simplest possible model, and in particular does not
allow for interactions between the planets, the GJ876 system (its
analogs considered here) is about as far as possible from the
idealized one variable pendulum model. Because of its highly
interactive nature, the GJ876 system analogs take a long time to
integrate and we consider ensembles of $\nens = 10^3$. 

At the other end of parameter space, we consider a model solar system
with relatively little interaction between the two planets. In this
case, we take the outer planet to be a Jupiter mass planet ($m_2 = 1
\, m_J$) in a 150 day orbit around a 1.0 $M_\odot$ star.  The second,
inner planet is taken to be a ``Super-Earth'' with mass $m = 0.01
m_J$.  The period of the inner planet is taken to be 75 days, and the
system is started in the 2:1 mean motion resonance. The orbital
eccentricities are $e$ = 0.1 for the inner planet and $e_2$ = 0.01 for
the outer planet.  This latter system is thus well approximated by the
circular restricted 3-body problem, which is used in the derivation of
the pendulum model for mean motion resonance (DM99).  We thus expect
ensembles of this class of solar system to follow the predictions of
the stochastic pendulum model. For these systems, the integrations can
be run faster and we consider ensembles of $\nens = 10^4$.

The effects of turbulence on mean motion resonance are illustrated by
Figures 1 -- 3. The first two of these figures show the fraction of
resonant states as a function of time for an ensemble of solar systems
that started out with the configuration of GJ876, and Figure 3
describes results for the less interactive Super-Earth systems.

\begin{figure} 
{\centerline{\epsscale{0.90} \plotone{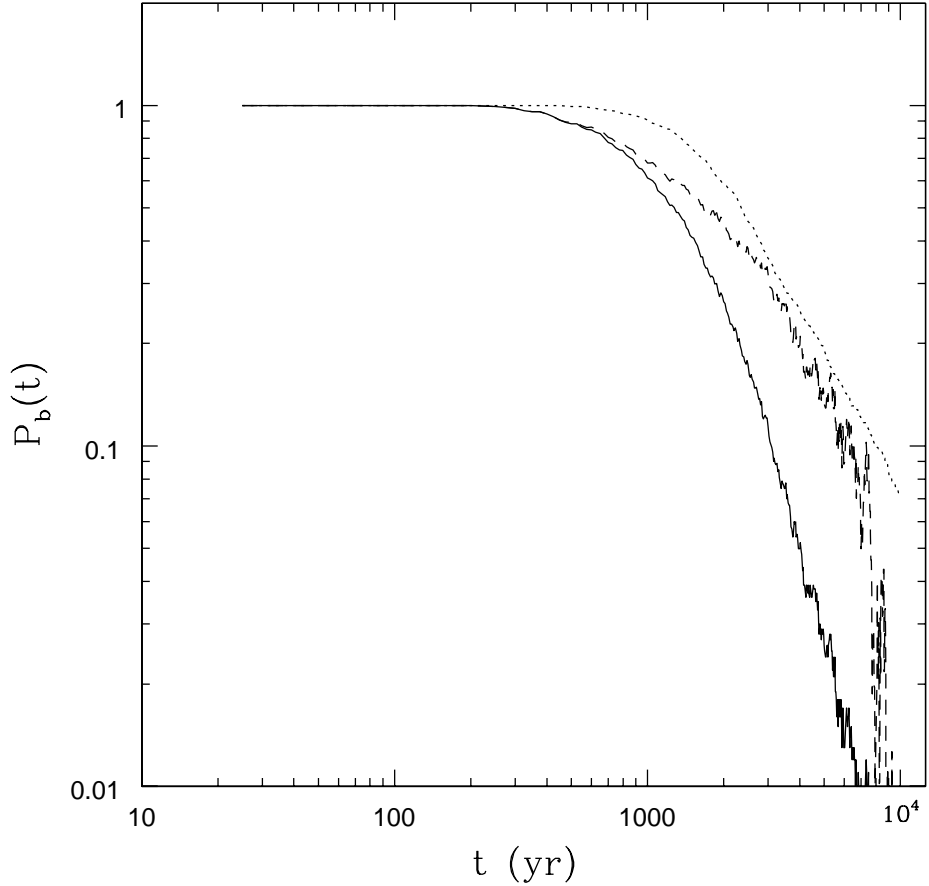} } } 
\figcaption{Results from numerical integrations of analogs of the 
GJ876 system to illustrate varying definitions of the bound fraction. 
Here the turbulent fluctuations have amplitude $(\Delta v)/v$ =
0.0002. The solid curve shows the ratio of the number of planets in
resonance to the number of planets in the original sample; the dashed
curve shows the ratio of the number of planets in resonance to the
number of planets that remain in orbit; the dotted curve shows the
fraction of planets that remain in orbit. }
\label{fig:defpbound}  
\end{figure}  

\begin{figure} 
{\centerline{\epsscale{0.90} \plotone{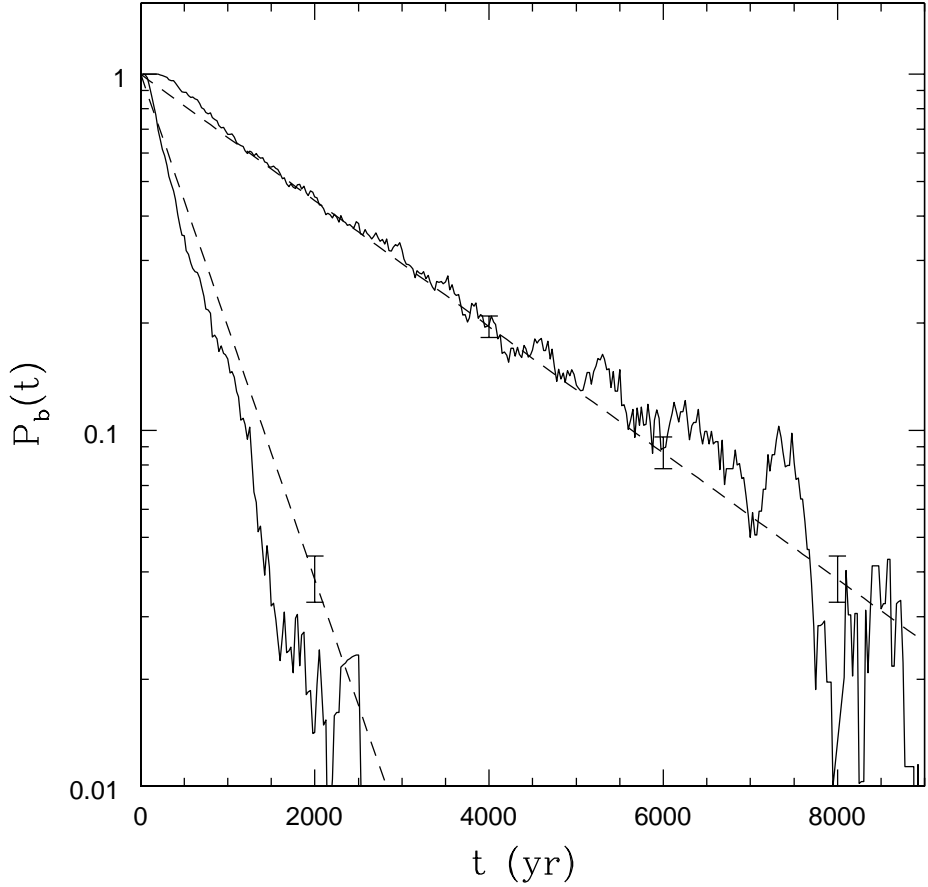} } } 
\figcaption{Time evolution of the bound fraction (number of planets in 
resonance divided by the number of planets that remain in orbit) for
numerical integrations of the GJ876 system. The two solid curves show
the numerical results obtained using two levels of turbulent
fluctuations, with $(\Delta v)/v$ = 0.0002 (right) and 0.0004 (left). 
Note that the diffusion constant $D$ sets the time scale for systems
to leave resonance, and that $D$ depends on the square of the
fluctuation amplitude, so the decay times for the two numerical
experiments should differ by a factor of four. For reference, the two
dashed lines show purely exponential decay with time scales that
differ by a factor of four. The error bars show the expected amplitude 
of root-$N$ fluctuations, which are somewhat smaller than the observed 
variations in the numerical results. } 
\label{fig:vamplitude} 
\end{figure}  

\begin{figure} 
{\centerline{\epsscale{0.90} \plotone{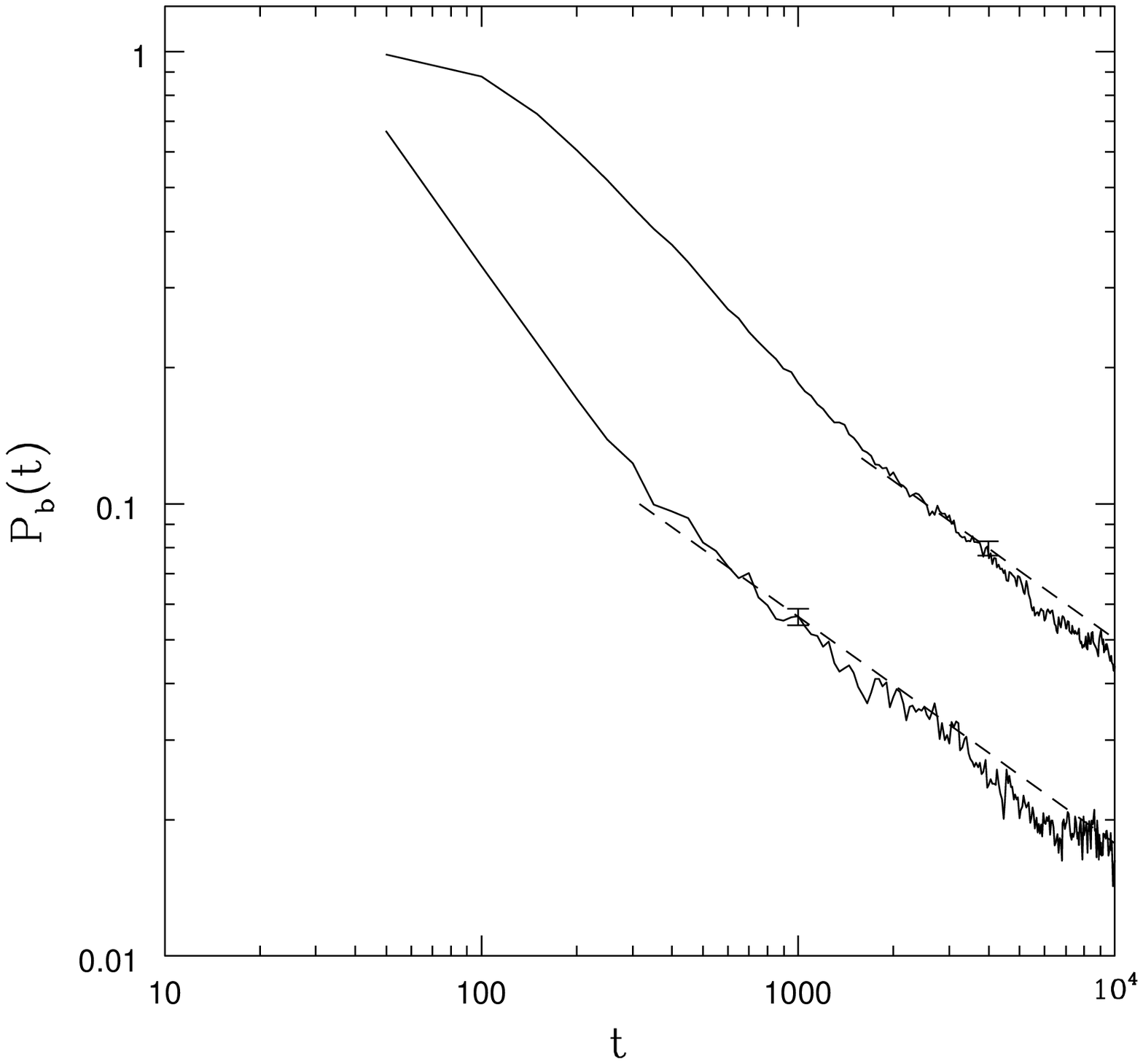} } } 
\figcaption{Time evolution of the bound fraction for numerical
integrations of the Super-Earth system, where $m_2 = m_J$, $P_2$ = 150
days, $m$ = 0.01 $m_J$, and $M_\ast = 1.0 M_\odot$. The bound fraction
is defined to be the number of planets in resonance divided by the
number of planets that remain in orbit.  The two solid curves show the
numerical results obtained using two levels of turbulent fluctuations,
with $(\Delta v)/v$ = 0.0001 (top) and 0.0002 (bottom). At late times,
when the bound fraction falls below $\pbd < 0.1$, the dashed lines
show the power-law behavior $\pbd \sim t^{-1/2}$ predicted by the
analytic model (which is valid in the absence of planet-planet
interactions).  The error bars show the expected amplitude of root-$N$
fluctuations. }
\label{fig:superearth} 
\end{figure}  

Figure 1 shows the various ways to account for the fraction of bound
states. The solid curve shows the ratio of the number of planets in
resonance to the number of planets in the original sample. However,
some planets are lost with time: systems that leave resonance often
experience orbit crossings, which can lead to planetary ejection. The
dotted curve shows the fraction of systems that have not ejected a
planet. Finally, the dashed curve shows the ratio of the number of
planets in resonance to the number of planets that remain in
orbit. For the remainder of this paper, we use this latter ratio
(number of systems in resonance over the number of surviving systems)
as the relevant fraction of bound states (this ratio is the most
directly observable).  Note that a substantial fraction of the systems
that leave resonance lose a planet. This ejection takes place because
of orbit crossing and subsequent planet-planet scattering.  Systems
are protected from such behavior while they remain in resonance, but
interactive systems (e.g., GJ876) readily lose planets in when
resonance is compromised.

Figure 2 shows how the time evolution of the bound fraction depends on
the level of turbulence. For an ensemble of GJ876 systems, the figure
shows the fraction of bound states as a function of time for two
levels of turbulent fluctuations, where $(\Delta v)/v$ = 0.0002 and
0.0004.  Note that in both cases, the bound fraction is nearly a
straight line on this log-linear plot, so that the time evolution is
nearly exponential with a well-defined decay rate, or, equivalently,
half-life (see \S 4 for further discussion). Notice also that the
diffusion constant $D$ sets the time scale for systems to leave
resonance, and that $D$ depends on the square of the fluctuation
amplitude, and hence the decay time scales differ by a factor of four.
The dashed lines shown in Figure 2 show a purely exponential decay,
with time scales that differ by exactly a factor of four. The
agreement shown in Figure 2 thus confirms our theoretical
expectations. The error bars plotted on the dashed lines show the
expected amplitudes of root-$N$ fluctuations, which are somewhat
smaller than the actual variations found in the numerical results; the
presence of planet-planet interactions (see also \S 4) increases the
level of these variations. 

Figure 3 shows the time evolution of the fraction of systems remaining
in resonance for an ensemble of Super-Earth systems. In this set of
experiments, two different levels of turbulence are used, so that the
amplitude of the velocity perturbations are $(\Delta v)/v$ = 0.0001
and 0.0002. This class of systems is close to the idealized, circular
restricted 3-body problem, and hence the behavior of the resonance is
expected to follow the predictions of the stochastic pendulum model.
As shown in Figure 3 (by dashed lines) the fraction of bound states
decreases as a power-law in time, with $\pbd \sim t^{-1/2}$ at late
times, roughly as predicted by \S 2.3. Given that the velocity
perturbations differ by a factor of two, one expects the coefficients
of the power-laws to also differ by a factor of two. However, the
numerical results show a somewhat wider gap, namely a factor of
$\sim$2.6. Although the size of the leading coefficient is correct 
in order of magnitude, the following three ambiguities arise in 
determining its predicted value from the analytic formulation: The
first issue is the mismatch between the simplified treatment of the
circular, restricted 3-body problem and full (18 variable)
integrations; as a result, the libration frequency of the numerical
system is not exactly the same as that given by the approximation of
equation (\ref{omegadef}).  The second difference is due to the
presence of a partial barrier; after systems leave resonance, they
have a somewhat lower probability $\rho_{\rm ret}$ of re-entering
resonance from an excited state because the physical system (with 18
phase space variables) is more complicated than the one-variable
pendulum (see Paper I for further discussion).  This effect should not
be overly severe for these Super-Earth systems, as they are close to
the regime of validity of the pendulum approximations (MD99), but the
partial barrier is still present. Yet another relevant process is that
planetary ejection can take place. In 3-body numerical integrations of
Earth-like planets with Jovian companions, when the periastron is the
same as in these systems, the expectation value of the ejection time
is about $3 \times 10^5$ orbits, or $6 \times 10^4$ yr (David et
al. 2003). In spite of these complications, however, the basic trends
shown by the numerical results are in qualitative (and rough
quantitative) agreement with the expectations of the simplest pendulum
model. Finally, we note that the error bars shown in Figure 3 show the 
amplitude of root-$N$ fluctuations, which are roughly the same amplitude 
as the variations seen in the numerical results.  
 
Now we can compare Figure 2 with Figure 3: For the interactive GJ876
systems of Figure 2, the fraction of bound states $\pbd$ is presented
as a log-linear plot, so that straight lines (as found in the
simulations) correspond to exponential decay.  For the Super-Earth
systems of Figure 3, however, the fraction $\pbd$ is presented as a
log-log plot, where straight lines (as found in these simulations)
correspond to power-law decay. Since exponential decay is much
stronger than power-law decay, we expect that highly interactive
systems will leave mean motion resonance much more readily than less
interactive systems. This effect is accounted for in the analytic
model developed \S 4. In particular, this model predicts power-law
decay of the fraction of bound states in the regime of minimal
interactions between planets, and exponential decay in the regime of
strong interactions.

\section{STOCHASTIC PENDULUM WITH DAMPING} 

This section considers the effects of damping on the maintenance of
mean motion resonance in the face of turbulent fluctuations.  We begin
with the dimensionless equation for a stochastic pendulum (see
equation [\ref{pendtwo}]), include the turbulent forcing term (see
equation [\ref{pendturb}] and Paper I), and then add a damping term,
\be
\frac{d^2\phi}{d\tau^2} + \gamma\frac{d\phi}{d\tau} + 
[1+\eta_k\delta([\tau]-\Delta\tau)] \sin\phi = 0 \, ,  
\label{pendamp} 
\ee
where $\gamma$ is the damping rate for the resonance angle. Here we
take $\gamma$ to be constant; its value is not well determined, but we
expect it to be small. For example, during planet migration, the
semimajor axes decrease with an effective ``damping rate'' $\gamma_a =
- {\dot a}/a$ and the disk tends to damp the eccentricity at a rate
$\gamma_e = - {\dot e}/e$. Since planet-planet interactions excite 
eccentricity during the migration epoch, the net damping rate $\gamma$
of the resonance is generally much smaller than either $\gamma_e$ or
$\gamma_a$.  As one example, even in the extreme case of eccentricity
damping where $\gamma_e = 100 \gamma_a$, as required to explain the
observed orbital elements of GJ876, the libration amplitude of the
resonance angle remains nearly constant, so that $\gamma \sim 0$ 
(see Figure 4 of Lee \& Peale 2002).

The basic equation of motion (\ref{pendamp}) can be converted into a
system of first order differential equations by introducing $V$, i.e.,
\be
\frac{d\phi}{d\tau}=V \, ,  \qquad 
\frac{dV}{d\tau}=-\gamma V-[1+\eta_k\delta([\tau]-\Delta\tau)]\sin\phi \, . 
\label{split} 
\ee
This set of equation corresponds to the following Fokker-Planck equation 
\be
{\partial P \over \partial \tau} = 
-V\frac{\partial P}{\partial\phi}+\gamma
\left( P+V\frac{\partial P}{\partial V}\right)+\sin\phi
{\partial P \over \partial V} + {D \over 2} \sin^2\phi
{\partial^2 P \over \partial V^2} \, , 
\ee
where the effective diffusion constant $D \equiv \langle \eta_k^2
\rangle /\Delta\tau$ determines the strength of the stochastic
forcing.  The terms that contain $\gamma$ encapsulate the effects of
damping. As the next step, we use the fact that $\phi$ changes more
rapidly than $V$, and hence we can average out the $\phi$ dependence
(see \S 2, Paper I, and MM04).  This averaging procedure removes the
$\sin\phi$ and $\partial/\partial\phi$ terms, and the Fokker-Planck
equation simplifies to the form
\be
\frac{\partial P}{\partial\tau}=\frac{D}{4}
\frac{\partial^2P}{\partial   V^2}+\gamma 
V\frac{\partial P}{\partial V}+\gamma P.
\ee
This equation can be considered as a ``modified'' diffusion equation.
Specifically, we can convert this equation into a diffusion equation
in standard form by making a suitable change of variables, i.e.,
\be
\tvar \equiv \frac{D}{8\gamma}\left(e^{2\gamma\tau}-1\right) \, , \qquad  
U \equiv Ve^{\gamma\tau} \, , \qquad 
P \equiv \pnew e^{\gamma\tau} \, .
\ee
With these definitions, we note that 
\be
{\partial P \over \partial\tau} = 
P \gamma + e^{\gamma\tau} {\partial \pnew \over \partial\tau}
= P\gamma + e^{\gamma\tau} \left( {\partial \pnew \over \partial \tvar} 
\frac{\partial \tvar}{\partial\tau} + \frac{\partial\pnew}{\partial U}
\frac{\partial U}{\partial \tau}\right)  =
P\gamma+e^{\gamma\tau}\left(\frac{D}{4}e^{2\gamma\tau}
\frac{\partial \pnew}{\partial \tvar}+\gamma U 
\frac{\partial \pnew}{\partial   U}\right) \, ,  
\ee 
and that 
\be
{\partial P \over \partial V} = e^{2\gamma\tau}
\frac{\partial\pnew}{\partial U}  \qquad {\rm and} \qquad 
\frac{\partial^2 P}{\partial V^2} = 
e^{3\gamma\tau}\frac{\partial^2\pnew}{\partial U^2} \, .
\ee
Using these results in the original equation, we find a diffusion 
equation of the form 
\be
P \gamma + e^{\gamma\tau}\gamma U
\frac{\partial \pnew}{\partial U} + 
\frac{D}{4}e^{3\gamma\tau}\frac{\partial\pnew}{\partial t}=
\frac{D}{4}e^{3\gamma\tau}\frac{\partial^2\pnew}{\partial U^2}
+e^{\gamma\tau}\gamma U\frac{\partial\pnew}{\partial U}+P\gamma  \, , 
\ee
which then reduces to the simpler form  
\be
\frac{\partial \pnew}{\partial \tvar}=
\frac{\partial^2\pnew}{\partial U^2} \, . 
\ee
This latter result is thus an ordinary diffusion equation.  The 
initial condition considered here is given by $P(\tau=0; V) =
\delta(V)$, so that all of the oscillators start in a bound (resonant)
state.  In terms of the new variables, this condition corresponds to
$\pnew(T=0; U) = \delta(U)$, and the solution for the distribution 
$\pnew(T,U)$ can be written in the form 
\be
\pnew (T, U) = \left( {1 \over \pi \tvar} \right)^{1/2} 
\exp\left(-\frac{U^2}{4 \tvar}\right)  =
\left[ {8 \gamma \over D (e^{2\gamma\tau} - 1) } \right]^{1/2} 
\exp\left[- \frac{2\gamma V^2 e^{2\gamma\tau}}{D\left(e^{2\gamma\tau}-1\right)}\right] \, . 
\ee
Finally, the distribution in terms of the original variables is given by 
\be
P (\tau, V) = 
\left[ {2 \gamma e^{2\gamma\tau} \over D \pi (e^{2\gamma\tau}-1) } \right]^{1/2}  
\exp\left[-\frac{2\gamma V^2 e^{2\gamma\tau}}{D\left(e^{2\gamma\tau}-1\right)}\right] \, .
\label{dampsolution}
\ee

Given the analytic form of equation (\ref{dampsolution}), we can now 
consider the asymptotic forms.  For the limit where $\gamma \tau \ll 1$, 
for small damping and/or short times, we can expand the exponentials 
to first order and find 
\be
P (\tau, V) = {1 \over (\pi D\tau)^{1/2} } 
\exp\left[ - {V^2 \over D\tau}\right] \, ,
\ee
which is the same result obtained earlier (see equation
[\ref{diffsolve}] and Paper I) with no damping.  In the opposite 
limit where $\gamma \tau \gg 1$, the distribution takes the form 
\be
P (\tau, V) = \left( {2\gamma \over D\pi} \right)^{1/2} 
\exp\left[ - {2\gamma V^2 \over D} \right] \, , 
\ee
i.e., the distribution approaches a gaussian form that is constant in time.

The general behavior described here can be understood through the
following heuristic argument: In the equation of motion (\ref{split})
for $V$, the time derivative has two contributions. For the stochastic
part, $V^2 \sim D \tau$ so that $V \sim \sqrt{D \tau}$ and hence
${\dot V} \sim V/2\tau$. For the damping contribution, ${\dot V} \sim
- \gamma V$. The two terms have opposite signs and are equal in
magnitude when $2 \gamma \tau$ = 1. For smaller values of $\gamma
\tau$, the stochastic term dominates and we recover the results of the
undamped stochastic oscillator. For larger values of $\gamma \tau$,
the damping term balances the stochastic term and the distribution
$P(V)$ becomes stationary.

Given the distributions derived above, the expectation value of 
$E = V^2/2$, the kinetic energy of the oscillators, is given by  
\be
\langle E \rangle = {1\over 2} \langle V^2 \rangle = 
{D \over 8 \gamma} {e^{2\gamma\tau}-1 \over e^{2\gamma\tau}} = 
{D \over 8 \gamma} \left[ 1 - e^{-2\gamma\tau} \right] \, .
\label{energyexp} 
\ee
In the limit of no (small) damping, the energy expectation value
becomes $\langle E \rangle \to D \tau / 4$, in agreement with the
result from the undamped case (Paper I). However, for any nonzero
value of $\gamma$, the expectation value approaches a constant
$\langle E \rangle \to D/(8\gamma)$ in the long time limit $\tau \to
\infty$. 

Numerical simulations of the stochastic pendulum, analogous to those
presented in Paper I, illustrate the behavior indicated by equation
(\ref{energyexp}). Specifically, Figure 4 shows the energy expectation
value for an ensemble of stochastic pendulums as a function of time.
The two solid curves show the result of numerically integrating the
stochastic pendulum equations (see also Paper I) for fixed diffusion
constant $D$ and two values of the damping rate $\gamma$ (here, in
dimensionless units, $D$ = 2 and $\gamma$ = 0.005, 0.05). The two
dashed curves show the prediction of equation (\ref{energyexp}) for
the same fluctuation amplitude and damping rates. The two treatments
are in excellent agreement. At early times, the energy expectation
value increases linearly with time, but then saturates and approaches
an asymptotic value $\langle E \rangle_\infty = D/(8\gamma)$ at late
times.

\begin{figure} 
{\centerline{\epsscale{0.90} \plotone{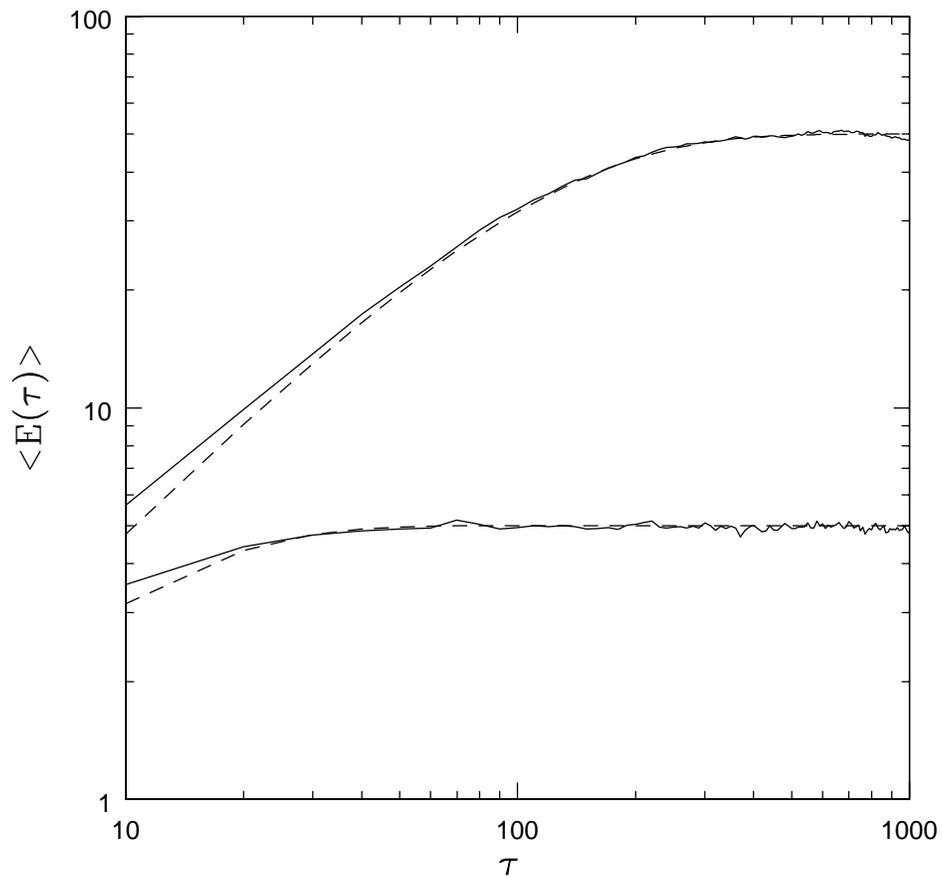} } } 
\figcaption{Energy expectation value for an ensemble of oscillators 
as a function of time. The two solid curves show the result of 
numerically integrating the stochastic pendulum equations for fixed
diffusion constant ($D$ = 2 in dimensionless units) and two values of
the damping rate ($\gamma$ = 0.005 and 0.05). The two dashed curves
show the predictions of equation (\ref{energyexp}). }
\label{fig:dampedenergy} 
\end{figure}  

Next we consider the bound fraction, i.e., the fraction of the systems
that remain in resonance as a function of time. For the sake of
definiteness, we define the bound fraction $\pbd$ to be the fraction of
states with $E = V^2 / 2 < 1$, i.e., we consider only the kinetic
energy in the accounting.  The bound fraction is thus given by the
integral
\be
\pbd (\tau) = \int_{-\sqrt{2}}^{\sqrt{2}}
\sqrt{\frac{2\gamma}{D\pi}}\frac{e^{\gamma\tau}}{\sqrt{e^{2\gamma\tau}-1}}
\exp\left[-\frac{2\gamma
  V^2 e^{2\gamma\tau}}{D
\left( e^{2\gamma\tau}-1\right)}\right] dV.
\ee
After defining new variables 
\be
z^2 = \frac{2\gamma V^2e^{2\gamma\tau}}{D\left(e^{2\gamma\tau}-1\right)} 
\qquad {\rm and} \qquad 
z_0^2 = \frac{4\gamma e^{2\gamma\tau}}{D\left(e^{2\gamma\tau}-1\right)},
\ee
this integral can be simplified to the form 
\be
\pbd (\tau) = {2 \over \sqrt{\pi}} \int_0^{z_0} e^{-z^2}dz = \erf(z_0) \, , 
\ee
where $\erf(z)$ is the error function (e.g., Abramowitz \&
Stegun 1970).  We are often interested in the limit where the argument
of the error function is small, so that we can expand $\erf (z)$
in a power series in $z$:
\be
\erf(z) = \frac{2}{\sqrt{\pi}}
\left(z - {z^3 \over 3} + {z^5 \over 10} - \ldots \right).
\ee
Keeping only the first order term, we can write the fraction 
of bound states in the form 
\be
\pbd (\tau) \approx 4 \left( {\gamma \over \pi D} \right)^{1/2} 
{e^{\gamma\tau} \over \left( e^{2\gamma\tau} - 1 \right)^{1/2} } \,  .
\label{boundgamma} 
\ee
This expression is valid in the limit $z_0 \ll 1$, which requires the 
following conditions to be met
\be
D \tau \gg 1 \qquad {\rm and} \qquad D \gg 4 \gamma \, . 
\label{conditions}
\ee
The first condition will be satisfied whenever turbulent fluctuations
persist for a long time, as expected in circumstellar disks associated
with young stars; this requirement is necessary for the bound fraction
$\pbd$ to be small in the absence of damping (Paper I).  The second
condition requires the diffusion constant to be larger than the 
damping rate. In other words, the size of the ratio $D/\gamma$ 
determines whether turbulence or damping dominates the dynamics. 

Now we can consider the limiting forms of the result from equation
(\ref{boundgamma}). In the limit of small damping or sufficiently 
short time scales, $\gamma \tau \ll 1$, we recover the prediction for 
the bound fraction from the undamped calculation of Paper I, i.e., 
\be
\lim_{\gamma \tau \to 0} \pbd = \left( {8 \over \pi D \tau } \right)^{1/2} \, . 
\label{pbold} 
\ee
In the opposite limit where $\gamma \tau \gg 1$, the bound fraction 
approaches an asymptotic (constant) value:  
\be
\lim_{\gamma \tau \to \infty} \pbd = 4 
\left( {\gamma \over \pi D} \right)^{1/2} \, . 
\label{pblimit} 
\ee
The asymptotic value of $\pbd$ will thus be small provided that the
damping rate $\gamma$ is small compared to the diffusion constant
(see equation [\ref{conditions}]). 

Figure 5 shows the results from integrations of two ensembles of
stochastic oscillators; the resulting behavior is in good agreement
with the analytic predictions derived above. This Figure shows the
fraction of bound resonant states as a function of time for the same
two ensembles of systems considered in Figure 4, i.e., stochastic
pendulums with turbulent fluctuations and two different values of the
damping rate $\gamma$.  As expected, the bound fraction initially
decreases like a power-law as indicated by in equation (\ref{pbold}),
where $\pbd (t) \sim t^{-1/2}$, and eventually saturates at the value
given by equation (\ref{pblimit}).

\begin{figure} 
{\centerline{\epsscale{0.90} \plotone{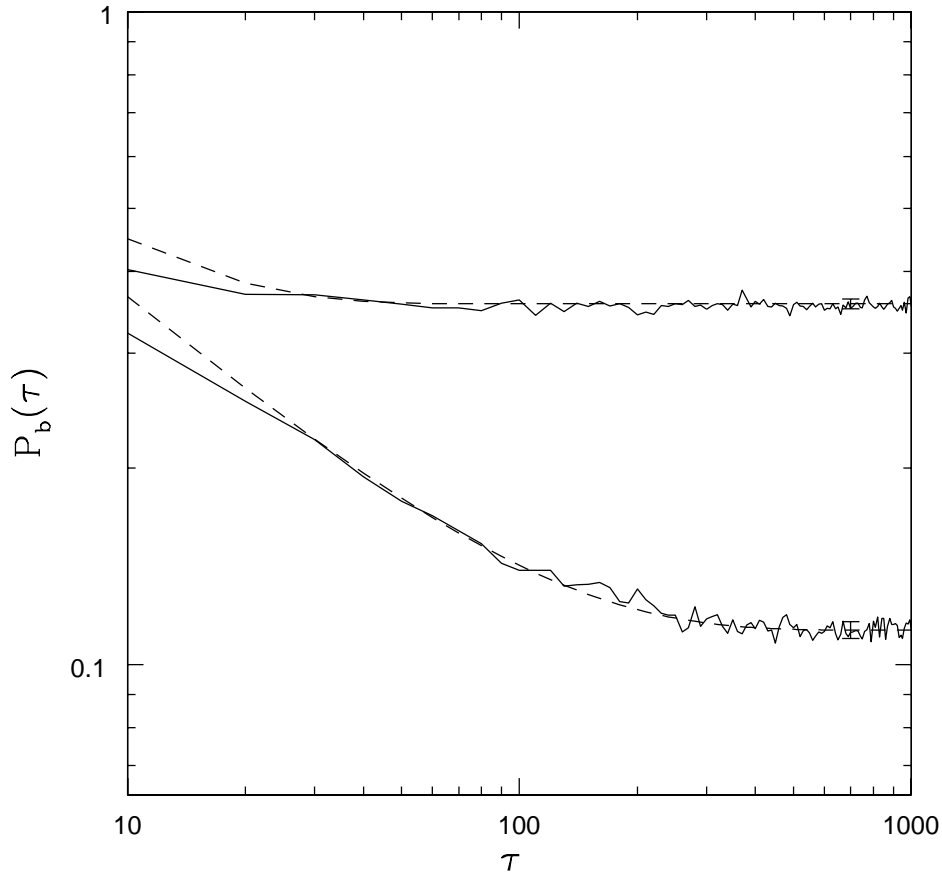}}}
\figcaption{Fraction of bound (resonant) states for an ensemble of
stochastic oscillators as a function of time. The two solid curves
show the result of numerically integrating the stochastic pendulum
equations for fixed diffusion constant ($D$ = 2 in dimensionless
units) and two values of the damping rate ($\gamma$ = 0.005 and
0.05). The two dashed curves show the predictions of equation
(\ref{boundgamma}). The error bars plotted at $\tau$ = 700 show 
the expected amplitudes of root-$N$ fluctuations (which roughly 
bracket the variations found in the numerical results). } 
\label{fig:dampedbf} 
\end{figure}

In order to estimate the size of this asymptotic value for $\pbd$, we 
rewrite the limiting expression (\ref{pblimit}) in the form 
\be
\pbd \approx {2 \over \etarms} \left( {N_\gamma \over \norb} \right)^{1/2} 
\approx 400 \left( {N_\gamma \over \norb} \right)^{1/2} \, , 
\ee
where $N_\gamma = \gamma \tau$ is the number of damping times and
$\norb$ is the number of orbits for which turbulence is active (see
also the definition of equation [\ref{diffusedef}]).  Unfortunately,
the relevant value of the damping time scale remains uncertain. The
time scale for planetary migration -- the time required for the
semimajor axis $a$ to change -- is typically a few Myr in these
systems, roughly comparable to the disk lifetimes.  In addition to
driving migration, the disk tends to damp eccentricity (for the most
part -- see Moorhead \& Adams 2008, Goldreich \& Sari 2003) on a time
scale that is roughly comparable to the migration time (e.g., Crida et
al. 2008). If we assume that turbulence is active over the same time
interval when damping is active, a few Myr, then $N_\gamma \sim 3$ and
the number of orbits $\norb \sim 10^7$.  These values would predict an
asymptotic value of the bound fraction $\pbd \sim 0.2$. However,
although eccentricity damping leads to a damping of the resonance
excitations, planet-planet interactions lead to eccentricity
excitation and tend to counteract its effects. As a result, values of
$\gamma$ and $N_\gamma$ are smaller than in this simple estimate, and
the asymptotic value of $\pbd$ will also be smaller. Further, all of
the relevant parameters (the size of the damping, the duty cycle of
the turbulence, the amplitude of the fluctuations, etc.) are expected
to vary from system to system. We thus expect damping effects to
counteract turbulence in some cases, but not in others. In any case,
the results of this section determine how the asymptotic value of
bound fraction $\pbd$ depends on the amplitude and duty cycle of the
turbulence, and the damping rate $\gamma$.

\section{RESONANCE MODEL INCLUDING INTERACTIONS} 

The analysis carried out thus far indicates that the fraction of
resonant states tends to decrease exponentially with time for highly
interactive systems (e.g., GJ876) in the presence of turbulence. On
the other hand, the fraction of bound states in less interactive
systems (e.g., the Super-Earth system) decreases as a power-law in
time, in agreement with the pendulum model of resonances (with
stochastic fluctuations). The pendulum model, with only one variable,
allows systems to freely random walk in and out of resonance; highly
interactive systems seem to have more difficulty re-entering a
resonant state. These results suggest that the barrier to returning
into resonance is due, in part, to interactions between the two
planets. These same interactions also lead to such systems leaving a
resonant state more easily in the presence of stochastic fluctuations.

In this section, we derive an equation for the evolution of the
resonance angle where interactions between the planets are included
(see also Holman \& Murray 1996, Quillen 2006).  As shown below, this
analysis initially retains many variables, and thus includes more
physics than the (one-variable) stochastic pendulum considered
previously. In order to obtain tractable results, however, we average
over many of the variables and find a Fokker-Planck equation for the
reduced and averaged problem. For the solution to this Fokker-Planck
equation, the fraction of bound (resonant) states decreases
exponentially with time when planet interactions are important; when
interaction terms are small, the solution leads to the same power-law
behavior $\pbd \propto \tau^{-1/2}$ found previously (see equation
[\ref{probtime}] and Paper I).

\subsection{Equations of Motion} 

In this section we derive the equations of motion to describe the
evolution of one particular resonance angle and the corresponding
orbital elements. Note that we cannot use the same resonance angle as
in \S 2. In that case, the simple pendulum equation resulted from the
circular restricted three-body problem (see MD99). In this context, we
want to allow for interactions between the planets (so we must move
beyond the restricted three-body problem) and hence the outer planet
can attain nonzero and time varying eccentricity (so the circular
approximation is no longer applicable). We thus consider the 
following (non-standard) resonance angle
\be
\phi= 2\lambda_2 - \lambda + 2\varpi_2 - 2\varpi \, ,
\label{newangle} 
\ee
where $\lambda$ is the mean longitude and $\varpi$ is the argument of
periastron. Note that this resonance angle is the analog (beyond the
circular, restricted 3-body approximation) of that from equation
(\ref{phidef}) for the case $|j_4|$ = 2.  Throughout this derivation,
the variables associated with the outer planet are denoted with the
subscript `2', whereas the variables associated with the inner planet
are left with no subscript.

For our chosen resonance angle (\ref{newangle}), the lowest order part 
of the disturbing function for the inner planet is given by 
\be
\langle \mathcal{R} \rangle =\frac{Gm_2}{a_2}
\left[ \mathcal{R}^{\mbox{(sec)}}_D+e^2e_2^2\left[ f_d(\alpha)+f_e(\alpha) 
\right]\cos\phi \right]
\ee 
and that for the outer planet is 
\be
\langle \mathcal{R}_2 \rangle = \frac{Gm}{a}
\left[ \alpha \mathcal{R}^{\mbox{(sec)}}_D+e^2e_2^2
\left[ \alpha f_d(\alpha)+f_i(\alpha) \right]\cos\phi \right] 
\ee 
where $e$ is eccentricity, $m$ the mass, ($a, a_2$) are the semimajor axes, 
and $G$ is the gravitational constant. The functions $f_d$, $f_e$, $f_i$  
are defined and discussed in MD99. To lowest order, the direct secular 
contribution is then given by 
\be
\mathcal{R}^{\mbox{(sec)}}_D=(e^2+e_2^2)f_{s,1}(\alpha) + 
e e_2 f_{s,2}(\alpha)\cos(\varpi_2-\varpi).
\ee
The corresponding equations of motion for the orbital elements of the inner 
planet then become 
\be
\dot{\Omega} =-\frac{3}{a^2}\frac{\partial \mathcal{R}}{\partial\lambda}=
-\frac{3Gm_2}{a^2a_2}e^2e_2^2\left[ f_d(\alpha)+f_e(\alpha)\right]\sin\phi, 
\ee
\be
\dot{e} =-\frac{1}{\Omega a^2e}\frac{\partial\mathcal{R}}{\partial\varpi} 
=-\frac{Gm_2}{\Omega a^2a_2e}\left[ee_2f_{s,2}(\alpha)\sin(\varpi_2-\varpi) +
2e^2e_2^2\left[f_d(\alpha)+f_e(\alpha)\right]\sin\phi\right], 
\ee
\be
\dot{\varpi} =\frac{1}{\Omega a^2e}\frac{\partial\mathcal{R}}{\partial e}= 
\frac{Gm_2}{\Omega a^2a_2e}\left[2ef_{s,1}(\alpha) + 
e_2f_{s,2}(\alpha)\cos(\varpi_2-\varpi)+2ee_2^2\left[f_d(\alpha)+
f_e(\alpha)\right]\cos\phi\right] \, .
\ee 
To complete the set, note that $\dot{\epsilon} = e^2 \dot{\varpi}/2$,
where $\epsilon$ is the mean longitude at epoch.  However, because
$\dot{\epsilon}$ is $e^2$ times smaller than $\dot{\varpi}$, we can
ignore time derivatives of $\epsilon$ in favor of time derivatives of
$\varpi$.  This assumption is valid within the range of validity for
these equations, which are the lowest order terms in an expansion in
the eccentricities ($e,e_2$).  The equations of motion for the orbital
elements of the outer planet are similar, i.e.,  
\be
\dot{\Omega}_2 = -\frac{3}{a_2^2}
\frac{\partial \mathcal{R}_2}{\partial\lambda_2}=
\frac{6Gm}{a_2^2a}e^2e_2^2
\left[ \alpha f_d(\alpha)+f_i(\alpha)\right]\sin\phi, 
\ee
\be
\dot{e}_2 = \frac{Gm}{\Omega_2 a_2^2ae_2} 
\left[ee_2 \alpha f_{s,2}(\alpha)\sin(\varpi_2-\varpi) + 
2e^2e_2^2\left[\alpha f_d(\alpha)+f_i(\alpha)\right]\sin\phi\right], 
\ee
\be
\dot{\varpi}_2 = 
\frac{Gm}{\Omega_2a_2^2ae_2}\left[2e_2 \alpha f_{s,1}(\alpha) + 
e \alpha f_{s,2}(\alpha)\cos(\varpi_2-\varpi)+2e^2e_2
\left[\alpha f_d(\alpha)+f_i(\alpha)\right]\cos\phi\right],
\ee

Next we rewrite $\lambda$ as $\lambda = \Omega t + \epsilon$, and assume
that $\dot{\Omega}t \ll \Omega$, and that $\ddot{\epsilon}\ll \ddot{\varpi}$.
The time evolution of the resonance angle is then given by
\be
\ddot{\phi}\approx 2\dot{\Omega_2} - 
\dot{\Omega}+2\ddot{\varpi_2}-2\ddot{\varpi} \, . 
\ee
Also, one can write $G M_\ast = \Omega^2 a^3 = \Omega_2^2 a_2^3$,
where $M_\ast$ is the mass of the central star.  Using this latter
result, we define 
\be
\cfun_r \equiv\frac{Gm_2}{\Omega a^2 a_2}
\left[f_d(\alpha)+f_e(\alpha)\right]=
\frac{m_2}{M_\ast} \Omega \alpha\left[f_d(\alpha)+f_e(\alpha)\right], 
\ee
\be
\cfun_{s1} \equiv\frac{Gm_2}{\Omega a^2a_2}f_{s,1}(\alpha)=
\frac{m_2}{M_\ast} \Omega \alpha f_{s,1}(\alpha) , 
\ee
\be
\cfun_{s2} \equiv \frac{Gm_2}{\Omega a^2a_2}
f_{s,2}(\alpha)=\frac{m_2}{M_\ast} \Omega \alpha f_{s,2}(\alpha),
\ee
and define $\cfuntwo_r,\cfuntwo_{s1},\cfuntwo_{s2}$ in an analogous 
way (where the superscript now refers to the second planet). With this 
choice of notation, the equations of motion take the form 
\be
\dot{\Omega} =-3\cfun_r \Omega e^2e_2^2\sin\phi, 
\ee
\be
\dot{e} =-\cfun_{s2}e_2\sin\theta-2\cfun_r ee_2^2\sin\phi, 
\ee
\be
\dot{\varpi} =2\cfun_{s1}+\cfun_{s2}\frac{e_2}{e}
\cos\theta+\cfun_r e_2^2\cos\phi, 
\ee
\be
\dot{\Omega}_2 =6\cfuntwo_r \Omega_2 e^2e_2^2\sin\phi, 
\ee
\be
\dot{e}_2 =\cfuntwo_{s2}e\sin\theta+2\cfuntwo_r e^2e_2\sin\phi, 
\ee
\be
\dot{\varpi}_2 =2\cfuntwo_{s1}+\cfuntwo_{s2}
\frac{e}{e_2}\cos\theta+\cfuntwo_r e^2\cos\phi,
\ee 
where $\theta\equiv\varpi_2-\varpi$.  

After taking a second time derivative of $\varpi$ and $\varpi_2$, we
can derive the equation of motion, which gives $\ddot{\phi}$ in terms
of the variables ($\phi,e,e_2,\Omega,\Omega_2,\theta$). For simplicity, 
we also leave the variables ($\dot{e},\dot{e}_2,\dot{\theta}$) in the
equation, but these can all be rewritten in terms of the variables 
listed above. The resulting equation of motion then takes the form
$$
\ddot{\phi} = 3e^2e_2^2
\left(4\cfuntwo_r \Omega_2 + \cfun_r \Omega\right) \sin\phi 
+ 2\left[\cfuntwo_{s2}\frac{\dot{e}}{e_2}\left(1 + 
\frac{\cfun_{s2}e_2^2}{\cfuntwo_{s2}e^2}\right) +
\cfun_{s2}\frac{\dot{e}_2}{e} \left( 1 + 
\frac{\cfuntwo_{s2}e^2}{\cfun_{s2}e_2^2}\right)\right]\cos\theta 
$$
\be
+ 2 \left(\cfuntwo_{s2}\frac{e}{e_2}-\cfun_{s2}\frac{e_2}{e}\right)
\dot{\theta} \sin\theta + 2\left(\cfuntwo_re\dot{e}-\cfun_re_2\dot{e}_2\right)
\cos\phi + 2\left(\cfuntwo_re^2-\cfun_re_2^2\right)\dot{\phi}\sin{\phi}.
\label{doublephi}
\ee
Consider the ordering of the terms in equation (\ref{doublephi}). 
All of the time derivative terms (i.e., those including $\dot{e}$ 
or $\dot{\phi}$), have an extra factor of $\cfun$, which introduces 
another factor of $\mu = m / M_\ast$. All of the terms in the equation 
are thus second order in $m/M_\ast$, except those coming from the 
$\dot{\Omega}$ and $\dot{\Omega}_2$ terms, which are only first order.
As a result, to leading order, we recover the usual pendulum equation
of motion to describe a mean motion resonance, i.e.,
\be
\ddot{\phi}= 3 e^2 e_2^2 
\left(4\cfuntwo_r \Omega_2 + \cfun_r \Omega \right) \sin\phi, 
\label{newpendulum}
\ee
where the effective frequency of the system is a linear combination of
the frequencies of the ``oscillators'' for each planet separately.
Notice that the frequency of this pendulum contains extra factors of
eccentricity compared to the simplified case considered in \S 2; this
difference arises because the two pendulum equations (\ref{pendone})
and (\ref{newpendulum}) correspond to different resonance angles.

\subsection{Inclusion of Turbulence} 

Now that we have derived the ``classic'' equation of motion $\phi$,
without fluctuations, we must include the effects of turbulence. In
this treatment, we incorporate turbulent forcing as a set of discrete
impulses in the equation of motion for the eccentricity, i.e.,
\be
\dot{e} = -\cfun_{s2}e_2\sin\theta-2\cfun_r ee_2^2\sin\phi + \xi 
\qquad {\rm where} \qquad \xi = \eta_k \delta 
\left( [t] - \Delta t \right) \, .
\label{eforcing}  
\ee
As shown above, the equations of motion for all of the variables
include eccentricity, and hence this ansatz for including turbulence
is convenient.  These impulses acting on the eccentricity then produce
corresponding impulses in the equation of motion for the resonance
angle $\phi$, and in the equations for ${\varpi}$ and ${\varpi_2}$.
Since the orbital elements of the external planet should not be
directly perturbed by an impulse acting on the inner planet, the
validity of including impulses in the latter equation is somewhat
subtle.  However, the effect shows up in the second derivative, which
acts like a force, which in turn changes when the inner planet
experiences a perturbation (here due to turbulence). Finally, we 
note that the distribution of impulses produced by equation 
(\ref{eforcing}) is more complicated than the uniform distribution of
velocity perturbations used in our numerical simulations. Because of
this complication, the amplitude of the eccentricity perturbations
will be different from the effective amplitude of the perturbations 
acting on the other variables (e.g., the velocity $V = {\dot \phi}$).

We can now write out the Fokker-Planck equation, which includes six
variables at this stage of derivation. In order to obtain tractable
results, we average over most of the oscillating variables. To start,
we note that the equation simplifies considerably after we average
over the angle $\phi$ itself. Although this approximation is often
used (e.g., MM04), its implementation in this case is somewhat
complicated. Here, the frequency of the largest oscillations of $\phi$
(those lowest order in $\cfun$) is proportional to $\sqrt{\cfun}$,
whereas the primary frequencies of the other quantities are
proportional to $\cfun$; note that $\cfun \ll 1$, so that $\phi$
oscillates with higher frequency. Further, the terms for which this
frequency difference does not hold are small and can be neglected.  
As a result, $\phi$ has effectively a higher frequency and can be
averaged out (see Paper I and MM04 for further justification).  
After averaging, the Fokker-Planck equation becomes
$$
\frac{\partial P}{\partial t} = 
\cfun_{s2}e_2\sin\theta\frac{\partial P}{\partial e}-
\cfuntwo_{s2}e\sin\theta\frac{\partial P}{\partial e_2} -
2 \left( \cfuntwo_{s1} - \cfun_{s1} \right) 
{\partial P \over \partial \theta}  
$$
$$
+\left(\cfuntwo_{s2}\frac{e}{e_2}P-\cfun_{s2}
\frac{e_2}{e}P\right)\sin\theta - 
\left( \cfuntwo_{s2} \frac{e}{e_2} - \cfun_{s2}\frac{e_2}{e} \right)
{\partial P \over \partial \theta} \cos\theta 
$$
$$
- 2\left(\cfuntwo_{s2}\frac{e}{e_2}-\cfun_{s2}
\frac{e_2}{e}\right)\dot{\theta}\sin(\varpi_2-\varpi)
\frac{\partial P}{\partial V} 
+ \frac{\diffuse_e}{2}\frac{\partial^2 P}{\partial e^2} 
+ 2 \diffuse_{ev} \cfuntwo_{s2}\frac{1}{e_2}
\left(1+\frac{\cfun_{s2}e_2^2}{\cfuntwo_{s2}e^2}\right)
\cos\theta\frac{\partial^2 P}{\partial V \partial e} 
$$
\be
+ 2 \diffuse_v \left\{ \left[ \cfuntwo_{s2}\frac{1}{e_2}
\left(1+\frac{\cfun_{s2}e_2^2}{\cfuntwo_{s2}e^2}\right)
\cos\theta \right]^2 + \frac{1}{2} \left(
\cfuntwo_r e \right)^2\right\} \frac{\partial^2 P}{\partial V^2} \, .
\label{fokkerpdist} 
\ee
Because of the difference in perturbation amplitudes for the
eccentricity $e$ and for the velocity $\dot \phi$ = $V$, the diffusion
constants $\diffuse_e$, $\diffuse_{ev}$, and $\diffuse_v$ are not
equal (in general).

\subsection{Apsidal Resonance} 

Next we consider how the angle $\theta$ evolves with time. We start
with the equation of motion for $\theta$ in the form
$$
\ddot{\theta}= \left[\frac{\dot{e}}{e_2}
\left(\cfuntwo_{s2}+ \cfun_{s2}
\frac{e_2^2}{e^2}\right)-\frac{\dot{e}_2}{e}
\left(\cfun_{s2}+ \cfuntwo_{s2}
\frac{e^2}{e_2^2}\right)\right]\cos\theta 
+ \left(\cfuntwo_{s2}\frac{e}{e_2}-\cfun_{s2}\frac{e_2}{e}\right)
\dot{\theta}\sin\theta 
$$
\be
+ 2\left(\cfuntwo_re\dot{e}+\cfun_re_2\dot{e}_2\right)
\cos\phi+2\left(\cfuntwo_re^2-\cfun_re_2^2\right)\dot{\phi}\sin\phi.
\ee
As discussed above, in this approximation we average over the (rapidly
oscillating) variable $\phi$. This averaging reduces the equation of
motion for $\theta$ to the form
\be
\ddot{\theta}= \left[\frac{\dot{e}}{e_2}
\left(\cfuntwo_{s2}+\cfun_{s2}\frac{e_2^2}{e^2}\right)-
\frac{\dot{e}_2}{e}\left(\cfun_{s2}+
\cfuntwo_{s2}\frac{e^2}{e_2^2}\right)\right]\cos\theta 
+\left(\cfuntwo_{s2}\frac{e}{e_2}-
\cfun_{s2}\frac{e_2}{e}\right)\dot{\theta}\sin\theta,
\ee
where now $\dot{e},\dot{e}_2,\dot{\theta}$ are $\phi$-averaged, 
and thus obey the reduced equations of motion 
\be
\dot{e} =-\cfun_{s2}e_2\sin\theta, 
\ee
\be
\dot{e}_2 =\cfuntwo_{s2}e\sin\theta, 
\ee
\be
\dot{\theta} = 2\cfuntwo_{s1}-2\cfun_{s1}+
\left(\cfuntwo_{s2}\frac{e}{e_2}-\cfun_{s2}\frac{e_2}{e}\right)\cos\theta.
\label{thetacancel}
\ee
When the masses of the planets are comparable, then $\cfun \sim
\cfuntwo$, and the first two terms on the right hand side of equation
(\ref{thetacancel}) nearly cancel. If, in addition, the planets have
the appropriate values of eccentricity $(e, e_2)$, then the final
terms can also nearly cancel out. The small remaining part of the
first two terms can then balance that of the second two terms,
$\dot{\theta} \approx 0$, and the system can be trapped in a
``$\theta$-resonance'' or apsidal resonance (for further discussion,
e.g., see Chiang \& Murray 2002, Murray-Clay \& Chiang 2006).
Numerical experiments show that the Super-Earth systems considered in
this paper are not in apsidal resonance, whereas the analogs of the
GJ876 system are.  Further, if the planetary masses in the GJ876
systems are changed slightly, then the value of $e/e_2$ adjusts so
that $\dot{\theta}\approx 0$.

To see how this latter adjustment works, consider  
the time derivative of $e/e_2$, i.e., 
\be
\frac{d}{dt}\left(\frac{e}{e_2}\right)=\frac{\dot{e}}{e_2}-
\frac{e\dot{e_2}}{e_2^2}=-
\cfun_{s2}\sin\theta-\cfuntwo_{s2}\frac{e^2}{e_2^2}\sin\theta.
\label{eratio} 
\ee
Keep in mind that both $\cfun_{s2}$ and $\cfuntwo_{s2}$ are negative.
Now, suppose that $\theta$ is increasing with $\sin\theta > 0$.  Then
$e/e_2$ would increase.  This behavior makes the positive term
$-\cfun_{s2} (e_2/e) \cos\theta$ in equation (\ref{thetacancel})
decrease in magnitude, and the negative term, $\cfuntwo_{s2} (e/e_2)
\cos\theta$, increase in magnitude, which would cause $\theta$ to
begin to decrease.  The same argument can be used to show that $e/e_2$
would decrease if $\theta$ became negative, which would result in
$\theta$ becoming more positive.

When the system is not in apsidal resonance, the variables 
($\theta,e,e_2$) all oscillate at a frequency given by a characteristic
value of $\dot{\theta}$.  As shown by equation (\ref{thetacancel}),
this characteristic value of $\dot{\theta}$ is of the order $\cfun$.
Likewise, when the system is in apsidal resonance, so that 
$\dot{\theta}\approx 0$, the equation of motion for $\theta$ reduces 
to the approximate form 
\be
\ddot{\theta} \approx -\left[\cfun_{s2}
\left(\cfuntwo_{s2}+\cfun_{s2}\frac{e_2^2}{e^2}\right)+
\cfuntwo_{s2}\left(\cfun_{s2}+\cfuntwo_{s2}
\frac{e^2}{e_2^2}\right)\right]\theta.
\ee
In this case, the variables $\theta$, $e$, and $e_2$ still oscillate
at a characteristic frequency given by $\cfun$.  This result justifies
averaging over the angle $\phi$, which has a characteristic
oscillation frequency of $\sqrt{\cfun}$ (keep in mind that
$\sqrt{\cfun} \gg \cfun$ because $\cfun \ll 1$).  However, we cannot
average the Fokker-Planck equation over $\theta$ (i.e., over $\varpi$
and $\varpi_2$) in either case, because $e$ and $e_2$ both vary on the
same time scale and because the amplitude of this variation is substantial. 

Next we consider the effect of turbulence on $\theta$.  As above, we will
add a turbulent forcing term to $\dot{e}$, 
\be
\dot{e}= -\cfun_{s2}e_2\sin\theta + \xi \, 
\ee 
where $\xi$ has the same meaning as before.  We can now write a system
of first order differential equations for $\theta,e,e_2$,
\be
\dot{\theta} \equiv \omega, 
\ee
\be
\dot{e}=-\cfun_{s2}e_2\sin\theta+\xi, 
\ee
\be
\dot{e}_2=\cfuntwo_{s2}e\sin\theta, 
\ee
$$
\dot{\omega}= - \left[\cfun_{s2}
\left(\cfuntwo_{s2}+\cfun_{s2}
\frac{e_2^2}{e^2}\right)+\cfuntwo_{s2}
\left(\cfun_{s2}+\cfuntwo_{s2}
\frac{e^2}{e_2^2}\right)\right]\cos\theta\sin\theta 
$$
\be
 + \left(\cfuntwo_{s2}\frac{e}{e_2}-
\cfun_{s2}\frac{e_2}{e}\right)
\left[2\cfuntwo_{s1}-2\cfun_{s1}+
\left(\cfuntwo_{s2}\frac{e}{e_2}-\cfun_{s2}
\frac{e_2}{e}\right)\cos\theta\right]\sin\theta 
 + \frac{\xi}{e_2}\left(\cfuntwo_{s2}+
\cfun_{s2}\frac{e_2^2}{e^2}\right)\cos\theta,
\ee
and find their equivalent Fokker-Planck equation:
$$
\frac{\partial Q}{\partial t} =  - \omega\frac{\partial Q}{\partial \theta}+
\cfun_{s2}e_2\sin\theta 
\frac{\partial Q}{\partial e}-\cfuntwo_{s2}e
\sin\theta\frac{\partial Q}{\partial e_2}+
\dot{\omega}\frac{\partial Q}{\partial \omega}+
\frac{\diffuse_e}{2}\frac{\partial^2 Q}{\partial e^2} 
$$
\be
+\frac{\diffuse_{e\omega}}{e_2}\left(\cfuntwo_{s2}+\cfun_{s2}
\frac{e_2^2}{e^2}\right)\cos\theta
\frac{\partial^2 Q}{\partial e\partial \omega}+
\frac{\diffuse_\omega}{2e_2^2}\left(\cfuntwo_{s2}+
\cfun_{s2}\frac{e_2^2}{e^2}\right)^2\cos^2(\theta)
\frac{\partial^2 Q}{\partial \omega^2} \, . 
\label{qdiffuse} 
\ee
Here we use $Q$ to denote the distribution function for the apsidal
resonance angle $\theta$, which is not to be confused with the
distribution function $P$ for the original resonance angle $\phi$.
This equation (\ref{qdiffuse}) has almost the same diffusion terms
(those containing the diffusion constants $\diffuse_e$,
$\diffuse_{e\omega}$, and $\diffuse_\omega$) as the Fokker-Planck
equation (\ref{fokkerpdist}) for $P$ (only the factors of 2 are
different).  This correspondence suggests that for systems starting in
apsidal resonance, the bound fraction for apsidal resonance should
show qualitatively the same time dependence as that of the
$\phi$-resonance. For example, we observe exponential decays in the
bound fraction for both angles in the GJ876 system (see also Paper I).
Furthermore, the effective diffusion constant for apsidal resonance is
approximately four times larger than the diffusion constant for
removing systems from $\phi$-resonance. Systems thus tend to leave
apsidal resonance four times faster, which is a trend seen in our
numerical simulations.

\subsection{Averaging the Fokker-Planck Equation} 

To make further process, we need to average over $\theta$.  This must
be done carefully because all the terms in the problem vary on the
same timescale as $\theta$.  We assume that the system is not in
$\theta$-resonance, and that $\theta=\omega t$, for some $\omega$.  As
a result, both $e$ and $e_2$ can vary with substantial amplitude about
their mean values (denoted here as $\ebar$ and $\ebar_2$) in the
following way:
\be
e =  \ebar + \frac{\cfun_{s2}\ebar_2}{\omega}\cos\theta 
\equiv \ebar + \delta e \cos\theta \, , 
\label{ecycle}
\ee
\be
e_2 = \ebar_2 - \frac{\cfuntwo_{s2}\ebar}{\omega}\cos\theta 
\equiv \ebar_2 - \delta e_2 \cos\theta \, , 
\label{etwocycle}
\ee
where the second two equalities define $\delta e$ and $\delta e_2$.
This ansatz assumes that these perturbations of the eccentricity are
smaller than the eccentricities $e$ and $e_2$. However, these
perturbations are nonetheless significant and, as derived below,  
provide the effects of interactions in this formulation. 

From the original equations of motion, we see that $\omega \approx
2\cfuntwo_{s1} - 2\cfun_{s1}$.  We can now average over $\theta$ in
equation (\ref{fokkerpdist}), and derive the simplified form for the
Fokker-Planck equation
$$
\frac{\partial P}{\partial t} =  
\frac{\diffuse_e}{2}\frac{\partial^2 P}{\partial e^2}+
2\diffuse_{ev} \left[\frac{\cfuntwo_{s2}}{\ebar_2}
\left(-\frac{\delta e_2}{2\ebar_2}\right)+
\frac{\cfun_{s2}\ebar_2}{\ebar^2}
\left(\frac{\delta e_2}{2\ebar_2}-
\frac{\delta{e}}{\ebar}\right)\right]
\frac{\partial^2 P}{\partial V \partial e} 
$$
\be
+ 2\diffuse_v \left\{ \frac{1}{2}\left[ \frac{\cfuntwo_{s2}}{\ebar_2}
\left(1+\frac{\cfun_{s2}\ebar_2^2}{\cfuntwo_{s2}\ebar^2}
\right)\right]^2 + \frac{1}{2}\left(\cfuntwo_r\ebar\right)^2\right\} 
\frac{\partial^2 P}{\partial V^2} \, ,
\label{fpeone} 
\ee
where only the first order terms in $\delta e$ and $\delta e_2$ have
been retained. In general, we can expand these formulae in a power
series in $\delta e/\ebar$ and keep higher order terms. For
simplicity, we use only the leading order terms here.  
When $e_2 \ll e$, we can use the result
\be
1 + \frac{\cfun_{s2}e_2^2}{\cfuntwo_{s2}e^2} \approx 1,
\label{approx} 
\ee 
and we can evaluate the integral (over $\theta$) exactly by 
contour integration.  In this case, we find 
\be
\frac{\partial P}{\partial t} = 
\frac{\diffuse_e}{2}\frac{\partial^2 P}{\partial e^2} +
2\diffuse_{ev} \cfuntwo_{s2} \left\{ 1 - \left[ 1 - 
\left( \delta e_2 / \ebar_2 \right) \right]^{-1/2} \right\} 
\frac{\partial^2 P}{\partial V \partial e} + 
2 \diffuse_v \left( \frac{\cfuntwo_{s2}}{2\ebar_2}\right)^2
{\partial^2 P \over \partial V^2} \, . 
\label{fpcontour} 
\ee
As a consistency check, note that if we reduce equation (\ref{fpeone})
using the approximation of equation (\ref{approx}) and then evaluate 
equation (\ref{fpcontour}) in the limit $\delta e_2/\ebar_2\ll 1$,
the resulting expressions agree.

At this point, it is advantageous to average over the eccentricity
$e$.  In the Fokker-Planck equation, the second derivative terms with
respect to $e$ suggest that the eccentricity is undergoing an ordinary
diffusion process, in which case its expectation value would grow like
$\sqrt{t}$.  In the true physical system, however, the eccentricity is
bounded both from below ($e \ge 0$) and from above ($e \le 1$). The
effective upper bound is even smaller because planets that reach a
sufficiently large eccentricity often experience orbit crossings and
eventual ejection. For sufficiently long time scales, the eccentricity 
is expected to approach a nearly stationary distribution, and it 
becomes a reasonable approximation to average over $e$.  We thus define 
the following quantities: 
\be
A \equiv \int \frac{\partial^2 P}{\partial e^2}\rho(e) de, 
\ee
\be
B \frac{\partial P}{\partial V} \equiv \int 
\frac{\partial^2 P}{\partial V\partial e}\rho(e) de,
\ee
where $\rho(e)$ is the distribution of $e$.  This approximation is 
straightforward when the original distribution $P(t,e,V)$ is a separable 
function; when this is not the case, the meaning of the approximation is  
more complicated. In either case, after averaging, the Fokker-Planck 
equation becomes 
$$
\frac{\partial P}{\partial t} =  
\frac{\diffuse_e}{2}A + 2\diffuse_{ev} \left[\frac{\cfuntwo_{s2}}{\ebar_2}
\left(-\frac{\delta e_2}{2\ebar_2}\right)+\frac{\cfun_{s2}\ebar_2}{\ebar^2}
\left(\frac{\delta e_2}{2\ebar_2}-\frac{\delta{e}}{\ebar}\right)\right]
B\frac{\partial P}{\partial V} \\
$$
\be
+ 2 \diffuse_v \left\{ \frac{1}{2}\left[ \frac{\cfuntwo_{s2}}{\ebar_2}
\left(1+\frac{\cfun_{s2}\ebar_2^2}{\cfuntwo_{s2}\ebar^2}\right)\right]^2
+ \frac{1}{2}\left(\cfuntwo_r\ebar\right)^2\right\} 
\frac{\partial^2 P}{\partial V^2} \, ,
\label{finalform} 
\ee 
which is now a one-dimensional diffusion equation that can be solved
analytically using standard methods.

\subsection{Solution and Results} 

Before proceeding further, we simplify notation by defining constants
($X,Y,Z$) so that the averaged version of the Fokker-Planck equation
(derived above) can be written in the form
\be
\frac{\partial P}{\partial t} = X + Y 
\frac{\partial P}{\partial V} + Z\frac{\partial^2 P}{\partial V^2} \, . 
\label{fpsimple} 
\ee
Using standard methods, we can write the corresponding solution in the form 
\be
P (t, V) = {1 \over 2\sqrt{\pi t Z}}
\exp\left[-\frac{\left(V-Yt\right)^2}{4Zt}+Xt\right] \, .
\label{fpsimplesolve} 
\ee
This solution corresponds to a distribution which is diffusing with
diffusion constant $Z$, drifting like $Yt$, and decaying on the
timescale $1/|X|$ (note that $X$ is expected to be negative to leading
order). In practice, however, we find that the quantity $X$ is small;
our numerical simulations indicate that $\diffuse_v / \diffuse_e \sim
10^4$ in typical cases. This ratio of diffusion constants is the same
as the ratio of the potential energy of the resonance to the potential
energy of the planetary orbit (equation [\ref{resenergy}]). The
diffusion constant $\diffuse_e$ measures the efficacy of turbulence in
changing the eccentricity, i.e., changing the orbit; the constant
$\diffuse_v$ determines how easily turbulence can change the speed $V
= {\dot \phi}$, which is easier to change by a factor of $\sim
10^4$. Because $\diffuse_e$ is small, the decay term in equation
(\ref{fpsimplesolve}) can be ignored on the intermediate time scales
relevant to this discussion, and we set $X = 0$ from this point
onward.

Given the solution for $P(t,V)$, we can integrate over the bound
states to find the fraction of systems in resonance as a function of
time, i.e.,
\be
\pbd (t) = \int_{-k}^k P(t,V) \, dV \, = {1 \over 2} 
\left[\erf\left(\frac{k+Yt}{2\sqrt{Zt}}\right)-
\erf\left(\frac{-k+Yt}{2\sqrt{Zt}}\right)\right],
\label{boundsix} 
\ee
where $k$ is the value of $V$ for which the oscillator begins
circulating.  When $k\gg Yt$, the distribution function is primarily
determined by diffusion, and the bound fraction takes the approximate
form
\be
\pbd (t) = \erf\left(\frac{k}{2\sqrt{Zt}}\right) \, ,
\label{interpower}
\ee 
as in the stochastic pendulum problem.  However, when $Yt\gg k$, at
late times we can approximate the error function with the asymptotic 
formula,
\be
\erf (x) \approx 1 - \frac{e^{-x^2}}{\sqrt{\pi}x} \, , 
\ee
which is valid in the limit $x \gg 1$ (Abramowitz \& Stegun 1970).  
In this limit, the fraction of bound states reduces to the form 
\be
\pbd (t) \approx  \left( {Z t \over \pi} \right)^{1/2} \, 
\exp\left(-\frac{Y^2}{4Z}t\right) \, 
\left[ {1 \over Yt - k} - {1 \over Yt + k} \right] 
\approx \left( {Z \over \pi} \right)^{1/2} \, 
{2 k \over Y^2 t^{3/2}} \, \exp\left(-\frac{Y^2}{4Z}t\right) \, .  
\label{boundsixlite}
\ee
The fraction of bound states thus displays exponential decay in this limit. 

In the limit where $e_2\ll e$, which holds for the GJ876 system, the
constants $Y$ and $Z$ reduce to the forms
\be
Y =-\diffuse_{ev} \, \frac{\cfuntwo_{s2}}{\ebar_2}
\left(\frac{\delta e_2}{\ebar_2}\right)B 
\qquad {\rm and} \qquad 
Z =\diffuse_v \left( {\cfuntwo_{s2}\over \ebar_2} \right)^2 \, , 
\ee
and hence the long term time evolution of the bound fraction is given by 
\be
\pbd \sim \exp\left[ - {B^2 \diffuse_{ev}^2 \over 4 \diffuse_v} 
\left(\frac{\delta e_2}{\ebar_2}\right)^2t\right] \, .
\label{interexp} 
\ee
From the definition of $\delta e_2$, we see that $\delta e_2/\ebar_2$
will be small when $\cfuntwo_{s2}/\cfun_{s1}$ is small,
i.e., when the system is not highly interactive.

We can now summarize the implications of this solution: If the system
is not sufficiently interactive, then the eccentricities do not vary
appreciably; in mathematical terms, the factors $\delta e_2$ and
$\delta e$ will be small, the parameter $Y$ will be small, and the
system will exhibit a power-law decrease in the fraction of bound
states as implied by equation (\ref{interpower}). In the opposite
limit of highly interactive systems, $\delta e_2$ and $\delta e$ are
nontrivial, $Y \ne 0$, and the solution displays the exponential decay
indicated by equation (\ref{interexp}) in the limit of long times. In
this interactive case, since the eccentricities are varying and
angular momentum is conserved, the semimajor axes of the planets will
vary. As a result, the mean motions ($\Omega, \Omega_2$) will also
vary and the location of the resonance will move around. This movement
of the resonance thus corresponds to the ``drift'' in $V$ obtained in
the solution to the Fokker-Planck equation in the above analysis.
When, in addition, the planets have large enough masses so that the
star itself moves substantially (as in the case of GJ876), the
location of the resonance moves even more.

\begin{figure} 
{\centerline{\epsscale{0.90} \plotone{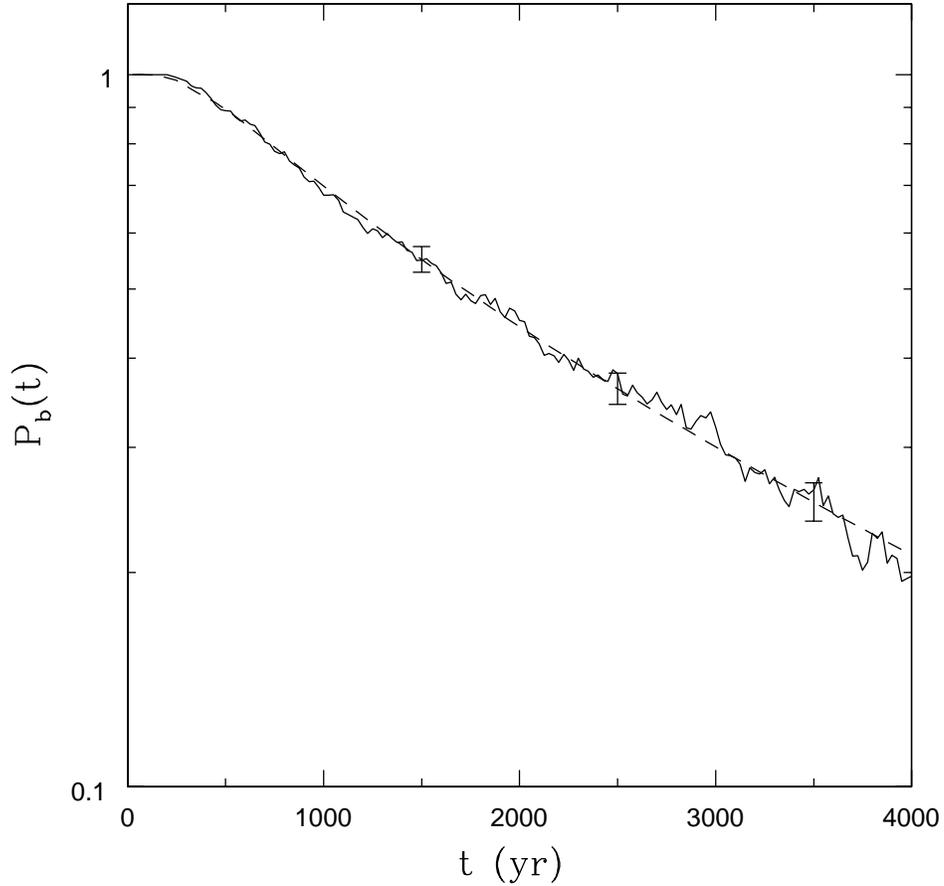} } } 
\figcaption{Comparison between numerical integrations of the GJ876 
system and the analytic results of this section. The solid curve shows
the numerical results for the fraction of bound states, i.e., the
number of planets in resonance divided by the number of planets in
orbit. The level of turbulence is given by $(\Delta v)/v$ = 0.0002.
The dashed curve shows the result expected from equation 
(\ref{boundsix}), where the parameters are adjusted to provide a
good fit. Note that the figure is presented as a log-linear plot, so
that a straight line corresponds to exponential decay. The error bars 
show the root-$N$ fluctuation amplitudes for comparison. } 
\label{fig:compare} 
\end{figure}

Next we want to compare the time scales for diffusion and drift.  In
the absence of the drifting effect, equation (\ref{interpower}) shows
that the bound fraction evolves on a time scale $\tdiff = k^2/(4Z)$.
Furthermore, since $\pbd \sim t^{-1/2}$, in order for turbulence to have
an effect, the evolution time must take place over many diffusion
times; for example, $\pbd \approx 0.1$ requires that $t \sim 100 \,
\tdiff$.  When the drift term becomes significant, equation
(\ref{boundsixlite}) shows that the bound fraction decays
exponentially with a time scale $\texp = 4Z/Y^2$.  To simplify this
discussion, let $e$ and $\mu$ represent the eccentricities and mass
ratios of both planets, and let $\delta$ = $\delta e_j/\ebar_j$ be 
the amplitude for eccentricity variations for both planets. The 
ratio of the two time scales is then given by 
\be
{\tdiff \over \texp } = \left( {k Y \over 4 Z} \right)^2 
\approx \, {\widetilde B}^2 \, \delta^2 \, {e^6 \over \mu} \, 
\left( {\diffuse_{ev} \over \diffuse_v} \right)^2 \, , 
\label{timeratio} 
\ee
where we have absorbed numerical factors of order unity into the
constant $\widetilde B$. For relatively large eccentricities, say, 
$e \sim 0.3$ near the peak of the observed distribution for extrasolar
planets, the factor ($e^6/\mu$) is of order unity (recall also that 
$e$ = 0.26 for GJ876). If the diffusion constants are comparable, then
equation (\ref{timeratio}) implies $\tdiff/\texp \sim \delta^2$. Thus, 
in order for exponential decay to be realized, the parameter $\delta$
must be relatively large, i.e., the changes in eccentricity must be
comparable to the mean values $\ebar$.  This condition is met when 
the two planets are relatively large and have comparable mass, so 
that $\cfun_{s2} \approx \cfuntwo_{s2}$ and $\delta \sim 1/2$ 
(see equations [\ref{ecycle}] and [\ref{etwocycle}]).  Because
exponential decay is much faster than power-law decay, this
requirement on $\delta$ is less severe then it might seem: Suppose,
for example, turbulence acts for 100 diffusion times, so that the
bound fraction $\pbd \approx 0.1$ in the absence of interactions. The
decay term is then roughly given by $\exp[-100 \tdiff / \texp] \sim
\exp[-100 \delta^2]$. Thus, the fraction of bound states can be
affected for more moderately interacting systems, e.g., with values of
$\delta \ge 0.1$. Keep in mind, however, that the ratio of time scales
depends on the other variables ($e, \mu, \diffuse_v, \diffuse_{ev}$)
and will thus vary greatly from system to system.

To test the predictions of this model, we compare the fraction of
bound states (resonant systems) from equation (\ref{boundsixlite})
with the fraction derived from an ensemble of numerical integrations
of the GJ876 system. The result is shown in Figure 6, which shows the
bound fraction -- the ratio of the number of systems in resonance to
the number of systems left in orbit -- as a function of time. Both the
analytic model and the numerical integrations show a nearly
exponential decline in the number of bound states (Figure 6 is
presented as a log-linear plot so that exponential decay corresponds
to a straight line). The analytic model is given in terms of error
functions with dimensionless arguments of the form $x = (k \pm Y t) /
2 \sqrt{Zt}$.  As a result, any fit to the numerical result
corresponds to a one parameter family of values of the parameters
$(k,Y,Z)$; the values will also depend on the choice of units.  For
the sake of definiteness, we have chosen $k$ = 1. Since $k$ is
approximately the libration frequency of the resonance, the other two
variables, here the diffusion constants $Y$ and $Z$ are then given in
terms of libration frequencies. In these units (and for time $t$
measured in years) we find that the values $Y \approx 5.3 \times
10^{-4}$ and $Z \approx 3.1 \times 10^{-4}$ provide a good fit (as
shown in Figure 6).

These results are sensible, at least in order of magnitude: From the
original diffusion equation, the diffusion constant $Z \sim
\omega_0^2/t$, where $\omega_0$ is the libration frequency and $t$ is
the diffusion time. The original stochastic differential equation also
defines the diffusion time, i.e., the time required for random
perturbations of size $\eta_k$ to change the libration speed by a total
displacement of order $\omega_0$. This time scale is given by $t \sim
9 P \omega_0^2 \Omega^{-2} [(\Delta v)/v]^2$, where $P$ is the period
of the inner planet (in years). Putting these two results together,
and inserting numerical values, we find that the diffusion constant is
given by $Z \sim 4 (9P)^{-1} \times 10^{-4} \sim 5 \times 10^{-4}$
yr$^{-1}$. Note that this value must be adjusted to units in which $k$
= 1 to compare with the fitting values of our analytic theory to the
numerical data; here, the libration period of GJ876 is about 9 years, 
so that $k$ = $2 \pi / P \sim 0.7$, which is close to unity, so that 
order of magnitude agreement obtains.

\begin{figure} 
{\centerline{\epsscale{0.90} \plotone{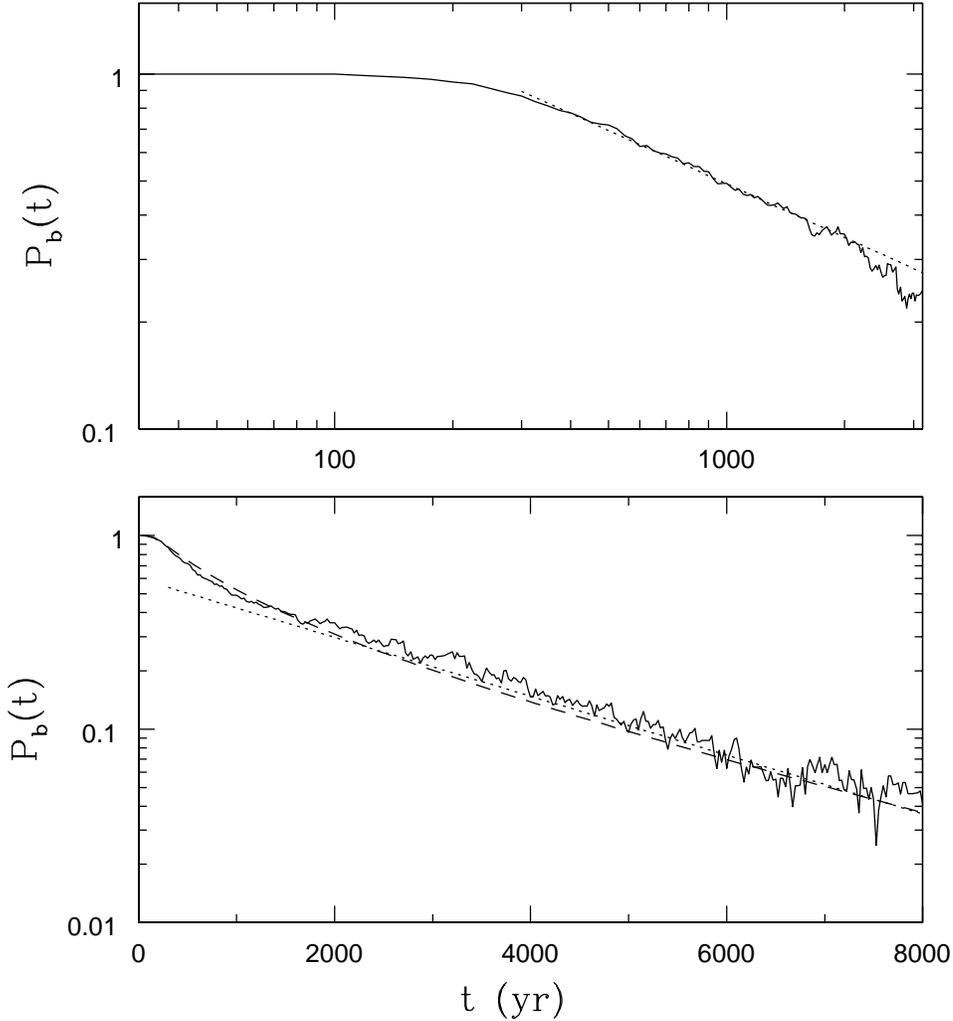} } } 
\figcaption{Results from numerical integrations of a less interactive 
version of the GJ876 system where the inner planet has 10 times
smaller mass than the observed system. The top panel shows the
fraction of bound resonant states as a function of time during the
early phase, when interactions have not yet had time to act and the
bound fraction $\pbd \sim t^{-1/2}$. The dotted line shows a power-law
fit to this result (the top panel is a log-log plot). The bottom panel
shows the longer term trend (as a log-linear plot), when the bound
fraction decreases exponentially with time; the dotted line shows a
fit to this result. The dashed curve (bottom panel) shows the result
expected from equation (\ref{boundsix}), where the parameters are
adjusted to provide a good fit; the results of this section thus
explain both the early power-law behavior and the later exponential
behavior with a single model. }
\label{fig:lightplanet} 
\end{figure}

One key result emerging from this analysis is that turbulence leads to
a power-law decrease in the fraction of resonant bound states in the
regime of little interaction between planets, whereas highly
interactive systems display exponential decay.  As a test of these
ideas, we have run the following numerical experiment: We begin with
an ensemble of GJ876 system analogs, but reduce the mass of the inner
planet by a factor of 10 in order to reduce the level of planet-planet
interactions. All of the other system parameters are kept the same as
before, and all of the systems are started in the 2:1 mean motion
resonance.  The level of turbulence, as set by the amplitude $(\Delta
v)/v$ of velocity perturbations, is also the same as before. 

The resulting time evolution for the fraction of bound states is shown
in Figure 7 for an ensemble of $\nens = 10^3$ systems.  The top panel
shows the early time evolution as a log-log plot, so that power-law
decay corresponds to a straight line. During this early phase of
evolution, planetary interactions (which are much weaker than in the
true GJ876 system) do not yet have time to act and cannot lead to
drift of the resonance location. As a result, the straight dotted line
in the top panel provides a good fit to the numerical results, with
the expected power-law slope of $-1/2$ (see \S 2 and Paper I). At
later times, the effects of planet-planet interactions gradually
accumulate, and their effect on the dynamics becomes important. As
shown above, at this stage, the time evolution of the fraction of
bound states turns over to an exponentially decreasing form. This
behavior is illustrated in the bottom panel of Figure 7, which is
presented as a log-linear plot, so that straight lines correspond to
exponential decay. Here, the dotted line shows a purely exponential
decay, which is expected for the late time behavior of these
systems. The dashed curve in the bottom panel shows the solution from
equation (\ref{boundsix}), where the parameters are adjusted to
provide a good fit. Note that the model developed herein smoothly
connects the non-interactive regime, with a power-law decrease in the
fraction of resonant states, with the highly interactive regime, with
an exponential decrease in the fraction of bound states.

In order to derive the simplified Fokker-Planck equation
(\ref{fpsimple}) and its solution for the time evolution of the
fraction of bound states (equation [\ref{boundsix}]), we have made a
number of approximations, and it is useful to summarize them here:
First, we averaged over the libration angle $\phi$ in the
Fokker-Planck equation; this approximation is well known and well
tested (e.g., MM04 and Paper I).  At the next stage of the
calculation, we adopted the ansatz of equations (\ref{ecycle}) and
(\ref{etwocycle}). This approximation captures the basic behavior of
the eccentricity variations for purposes of modeling the diffusion of
the energy of the resonance; if we needed the full time dependence of
the eccentricities, however, their full equations of motion should be
retained. With this simplified description for the eccentricities, we
then averaged over the apsidal angle $\theta = \varpi_2 - \varpi$, and
thereby obtained a diffusion equation in two variables ($e$ and $V$).
Although the resulting two dimensional diffusion equation can be
solved, it is rather cumbersome. In addition, the eccentricity values
are not free to diffuse to arbitrary values, but rather are confined
to the range $0 < e < e_{\rm max}$, where planets with $e > e_{\rm max}$ 
tend to experience orbit crossing and ejection. We thus average the 
diffusion equation over the eccentricity to obtain the simplified form
of equation (\ref{finalform}).  Although this procedure represents an
uncontrolled approximation, and hence this last averaging is the least
justified, the uncertainties can be encapsulated in the constants $A$
and $B$.  The solutions to the resulting diffusion equation can then
be found exactly, and they contain additional behavior, beyond that of
the solutions found in \S 2.  In particular, these solutions show that
the fraction of bound states can decrease exponentially when the
interactions between planets are significant, and this trend agrees
with the results of the corresponding numerical simulations of the
full 3-body problem. Finally, we note that equation (\ref{fpsimple})
is more robust than its derivation -- this form of the Fokker-Planck
equation represents a straightforward generalization of that obtained
earlier for the simpler stochastic pendulum.
 
\section{CONCLUSION} 

\subsection{Summary of Results} 

Building on earlier work (Paper I), this paper explores the effects of
stochastic fluctuations on the maintenance of mean motion resonance in
planetary systems. The principal result of this study is that
turbulence can readily compromise mean motion resonance for the
fluctuation amplitudes predicted by MHD simulations and for the solar
system architectures observed in extrasolar planetary systems.
Furthermore, we can understand the physical processes involved using a
(primarily) analytic approach.  A more specific outline of our results
is given below:

The most important quantity calculated in this paper is the fraction
of systems that remain in resonance as a function of time. This
fraction of bound states $\pbd(t)$ is defined to be the number of
systems in resonance divided by the number of systems that remain
intact (without ejecting a planet). This distinction is necessary
because mean motion resonance protects multiple planet systems from
instability, so that planetary ejection takes place with significant
probability after a mean motion resonance is compromised. As a result,
the fraction of systems remaining in resonance depends on whether the
fraction is calculated relative to the original number of systems in
the ensemble or the number of multiple planet systems remaining
(Figure 1).

The effects of stochastic fluctuations on mean motion resonance act in
qualitatively different ways in systems that are highly interactive
and those which are not (compare with Figure 2 with Figure 3).  In
systems with relatively little interaction between the planets, the
fraction of bound states (systems in resonance) decreases with time as
a power-law, specifically $\pbd \sim t^{-1/2}$.  Systems that display
this type of behavior are close to the idealized, circular restricted
3-body problem; the example considered in this paper contains a
Jupiter in a nearly circular orbit (initially) in resonance with a
Super-Earth on an interior orbit.  The simple pendulum model of mean
motion resonance (MD99) with turbulent forcing (Paper I) also derives
from the circular restricted 3-body problem and leads to this same
power-law time dependence (see equation [\ref{probtime}]). For these
non-interactive systems, ensembles of the stochastic pendulum (\S 2.1
and 2.2), solutions to the Fokker-Planck equation (\S 2.3), and direct
numerical integrations of the 3-body problem (\S 2.4) all predict the
same time dependence for the fraction of systems remaining in mean
motion resonance. In addition, the departures of the numerical results
from the expectation values are well-characterized by the amplitude of
root-$N$ fluctuations (Figure 3).  For this regime, we can summarize
these results by writing the expected fraction of surviving resonances
in the form $\pbd \approx C / \norb^{1/2}$, where $\norb$ is the total
number of orbits for which turbulence is active, and where the
dimensionless constant $C$ depends on the amplitude of the
fluctuations and the probability $\rho_{\rm ret}$ of returning to
resonance after leaving (we expect $C \sim 10 - 100$; see Paper I).
Thus, for $\norb \approx 10^6$, we expect $\pbd \approx 0.01 - 0.1$.

In highly interactive systems, the fraction of bound states (systems
in resonance) decreases more rapidly with time (see Figure 2) and
eventually shows an exponential decay.  In this paper, we have
developed a generalized treatment for mean motion resonance that
includes planetary interactions in the model equations (\S 4). This
treatment initially retains six variables, and thus contains more
physics than the simple one-variable pendulum model; in particular, we
include terms that describe the interaction between planets due to
their mutual excitation of eccentricities.  In this model, the
fraction of bound states $\pbd (t)$ shows two limiting regimes of
behavior: For the case of minimal planet-planet interactions, or for
sufficiently short time scales while the diffusion effects dominate,
the bound fraction shows power-law behavior $\pbd \sim t^{-1/2}$. For
highly interactive systems, or for sufficiently late times, the
evolution of the distribution function is dominated by drift terms due
to interactions, and the model predicts an exponential decrease in the
fraction of bound states (equation [\ref{interexp}]). This predicted
exponential behavior is in good agreement with that indicated by
numerical experiments of highly interactive systems such as GJ876 (see
Figures 2 and 6). For these systems, the variations in the numerical
results for the bound fraction $\pbd (t)$ are somewhat larger than
expected from root-$N$ fluctuations alone; this complication is most
likely due to the planetary interactions, which allow for additional
degrees of freedom.  

Systems that contain moderate levels of planet-planet interactions can
fall in an intermediate regime, where the fraction of bound states
initially decreases as a power-law with $\pbd \sim t^{-1/2}$ for a
substantial time interval, but eventually switches over to an
exponential decay (see Figure 7). Furthermore, the duration of the
initial power-law phase depends on the level of interactions in the
system.  During the early time evolution, interactions have little
effect, and the system acts essentially like the simple stochastic
pendulum; at later times the effects of interactions accumulate, and
the fraction of bound states decays exponentially. These two regimes
are connected at intermediate times, when $\pbd(t)$ behaves as a
steeper power-law in time (see equation [\ref{boundsixlite}]). The
model developed in \S 4 accounts for this early power-law behavior,
the later exponential decay, and the smooth matching between the
regimes.

Finally, for completeness, we have also studied the ramifications of
including a damping term in the model equations for mean motion
resonance (\S 3). For example, this damping can be driven by torques
from the circumstellar disk that is responsible for planetary
migration.  In this case, for sufficiently short time scales, the
distribution function for the ensemble of states evolves like that of
the stochastic pendulum; the energy expectation value increases
linearly with time (equation [\ref{energyexp}]) and the fraction of
bound states decreases as a power-law $\pbd \sim t^{-1/2}$ (equation
[\ref{pbold}]). At later times when $\gamma t \gg 1$, however, both
the energy expectation value (equation [\ref{energyexp}]) and the
fraction of bound states (equation [\ref{pblimit}]) approach
asymptotic values. The analytic solutions to the Fokker-Planck
equation (\S 3) are in excellent agreement with results obtained from
numerical integrations of the stochastic pendulum equation with
damping (see Figures 4 and 5).

\subsection{Discussion} 

This paper extends the analysis of Paper I and bolsters its main
conclusion, i.e., that turbulence implies mean motion resonances in
extrasolar planetary systems should be relatively rare (roughly at the
level of a few percent), unless turbulence has a limited duty cycle.
The fact that turbulence is capable of removing systems from resonance
is not surprising. As shown in equation (\ref{resenergy}), the
potential energy associated with a bound resonant state is much
smaller (typically, by a factor of $\sim 10^4$) than the binding
energy of the planet within the gravitational potential well of the
star.  Since planets can be ejected with relative efficiency during
the early phases of solar system evolution (e.g., Rasio \& Ford 1996),
it makes sense that additional perturbations (here, turbulent
fluctuations) can remove systems from resonance.

Given that the results of this paper (and Paper I) suggest that mean
motion resonances should be rare, it is useful to compare this
prediction to existing observational data. However, we first note that
a detailed comparison is premature, due to the small number of systems
found to date, and due to selection effects.  In the current sample,
30 systems are observed to have multiple planets. As outlined in the
introduction, one system (GJ876) is found to be deep in a 2:1
resonance, whereas three other systems are either in resonance or
close.  If all three of these latter systems are actually in
resonance, one ``estimate'' for the nominal rate of resonances would
be 4/30 or 13\%; if only GJ876 is truly in resonance, this rate
becomes 1/30 or 3\%.  These estimates suggest that mean motion
resonances are at least {\it somewhat} rare.  Besides the small
numbers involved, this percentage is uncertain for several reasons:
One important issue is that the fraction of systems in mean motion
resonance calculated here is the survival rate, i.e., the fraction of
systems that start in resonance and are not removed via turbulence. A
large fraction of the systems that leave resonance eject one of the
planets, and hence would not be included in the number of observed
multiple planet systems.  Acting in the other direction, another
consideration is that not all of the multiple planet systems in the
current sample were ever (necessarily) in resonance.  In addition, the
observations are not complete, so that many of the single planet
systems in the current sample might have companions (perhaps with
nearly integer period ratios), which could either add to the number of
resonant systems or add to the number that are not in resonance. Our
understanding of these issues will benefit from future observations
and hence better statistics.

This work also shows that for systems that leave resonance, experience
orbit crossing, and eject a planet, the surviving planet typically
displays an eccentric orbit. Although a detailed statistical
description of the resulting distribution is beyond the scope of this
paper, we note that the full range of possible eccentricities is
realized. This finding is consistent with previous simulations of
multiple planet systems undergoing inward migration; in this setting,
one of the planets is often scattered out of the system, and the
remaining planets (for an ensemble of systems) are left with semimajor
axes and eccentricities that fill the $(a,e)$ plane in a manner
consistent with the current observational sample (Adams \& Laughlin
2003, Moorhead \& Adams 2005). Preliminary numerical experiments of
this process including turbulent fluctuations (Moorhead 2008) indicate
that the orbital elements of the surviving planets continue to fill
the $(a,e)$ plane, but better statistics are needed (both from the
simulations and for the observational sample). In any case, 
planetary scattering and ejection --- perhaps enhanced by turbulence
removing systems from resonance --- provides one viable mechanism to
increase the orbital eccentricity of extrasolar planets.

Extrasolar planetary systems display a great deal of diversity, and
the effects of turbulence will be different for varying solar system
architectures.  The results of this paper show that highly interactive
systems (like GJ876) can reach a state of evolution where the fraction
of resonant systems decays exponentially. Highly interactive systems
are thus the least likely to be able to remain in resonance. This
finding, however, makes the existence of GJ876 itself, which is
observed to be deep in resonance, an even greater enigma. One way to
account for the existence of the 2:1 resonance in the GJ876 system is
for its disk to have had low mass (to reduce the level of turbulent
fluctuations) and/or a short survival time after planet formation (to
reduce the duty cycle of the effect). However, this explanation for
the survival of the resonance would make it more difficult for the
disk to produce relatively large planets.  Recall that both
theoretical considerations (Laughlin et al. 2004) and observational
findings (e.g., Johnson et al. 2007) suggest that M stars have
difficulty forming planetary companions with Jovian masses.

The results of \S 4 suggest that systems with the smallest levels of
interaction between planets would have the best chance of surviving in
a resonant state. However, this hypothesis is not entirely true: For a
given orbital spacing, for example that corresponding to a 2:1 period
ratio, the level of interactions decreases as the planetary masses
decrease.  On the other hand, planets with the smallest masses produce
the smallest gaps in their circumstellar disks, and hence experience
greater levels of turbulent forcing; in other words, the disk
reduction factor $\Gamma_R$ due to gaps is smaller for low mass
planets (Paper I). Planetary systems with only lower mass planets,
like the newly discovered HD 40307 system with masses of $m_P$ = 4 --
9 $M_E$ (Mayor et al. 2008), may experience greater levels of
turbulent forcing than systems with larger planets, and hence would be
more likely to be removed from resonance. Taken together, these
theoretical results to date suggest that the systems with the best
chances of maintaining mean motion resonance in the face of turbulence
are those with planets of moderate mass and nearly circular orbits.
The ``Super-Earth'' system of this paper --- with a Jovian planet in
an outer, nearly circular orbit and a smaller inner planet ---
provides one good example (see Figure 3).

The numerical simulations indicate another difference between the
highly interactive and less interactive systems. After leaving a
resonant state, the Super-Earth systems take a long time for their
planets to experience orbit crossing and eventual planetary ejection.
The GJ876-type systems eject their planets much more readily. As a
result, the less interactive systems (e.g., Super-Earth systems) can
stay ``close'' to resonance for a long time, where the period ratios
are close to 2:1, but the other orbital elements do not have the
proper values to be in true resonance.

This paper has made progress on our understanding of how well the
simplest pendulum equation works as a model for mean motion
resonances, including stochastic forcing terms, and we have
generalized the model to include damping (\S 3) and planetary
interactions in an approximate manner (\S 4). Nonetheless, a great
deal of work remains to be done. One important issue is to understand
how turbulence acts in the earliest stages of planet migration, i.e.,
when the planets are formed and first become locked into resonance.
This issue requires much more extensive numerical work and is thus
beyond the scope of this paper. Recent work (see Moorhead 2008) is
starting to study how turbulence affects the early phases of
migration. The issue of turbulence during planetary formation is also
being considered elsewhere (e.g., Masahiro et al. 2007). In addition,
we have not included the back reaction of how planets (e.g., through
gap clearing) can affect the generation of turbulent fluctuations
through the magneto-rotational instability. All of these issues
provide fruitful avenues for future research.

In the coming years, as the statistics of observed multiple planet
systems become sufficiently complete, the fraction of systems in
resonance will provide important constraints on their formation and
subsequent evolution. Since mean motion resonance can be compromised
relatively easily, as shown herein, planetary systems observed in such
bound states must have followed restricted historical paths (e.g.,
with little interaction and/or large dissipation). This type of
analysis has been considered for the resonances in our Solar System
(Peale 1976, Malhotra 1993) and for the GJ876 system (e.g., Lee \&
Peale 2002), and will soon provide interesting constraints on other
extrasolar planetary systems. However, the greatest wealth of
information on this topic will come when the observed sample of
multiple planet systems is large enough to determine the fraction of
actual mean motion resonances and the fraction of near-resonances.

$\,$

\acknowledgements
This paper was the result of a summer research project for DL at the
University of Michigan. We thank Jeff Druce, Jake Ketchum, Greg
Laughlin, Renu Malhotra, Althea Moorhead, and Ellen Zweibel for many
useful discussions and suggestions.  This work was supported in part
by the Michigan Center for Theoretical Physics. FCA is supported by
NASA through the Origins of Solar Systems Program via grant
NNX07AP17G.  AMB is supported by the NSF through grants CMS-0408542
and DMS-604307. In addition, AMB and FCA are jointly supported by
Grant Number DMS 0806756 from the NSF Division of Applied Mathematics.

\end{document}